\documentclass[]{onecat}
\usepackage{caption} 
\usepackage{hyperref}
\usepackage{cleveref}
\usepackage{verbatim}
\usepackage{amssymb} 
\usepackage{graphicx}
\usepackage{floatrow}
\usepackage{subcaption}
\usepackage{listings}
\usepackage{subcaption}
\usepackage{algorithm}
\usepackage{graphicx}
\usepackage{floatrow}
\usepackage{makecell}
\usepackage{enumitem}
\usepackage{wrapfig}  

\usepackage{subcaption} %

\usepackage[toc,page,header]{appendix}

\usepackage{pifont}

\definecolor{sh_blue}{rgb}{0,0.60,0.93}
\definecolor{sh_gray2}{rgb}{1,0.89,0.75}
\definecolor{color3}{rgb}{0.95,0.95,0.95}
\definecolor{title}{rgb}{0.75,0.51,0.96}
\definecolor{mygray}{gray}{.9}
\definecolor{bluegreen}{rgb}{0.44, 0.64, 0.77}
\definecolor{gray_venue}{rgb}{0.53,0.52,0.52}
\definecolor{color5}{rgb}{1,0.96,0.88}

\usepackage{tcolorbox}
\tcbuselibrary{skins, listingsutf8} 
\tcbset{
    examplebox/.style={
        colback=green!5!white, 
        colframe=green!75!black, 
        coltitle=green!50!black, 
        fonttitle=\bfseries, 
        boxrule=0.5mm, 
        sharp corners, 
        enhanced, 
        attach boxed title to top left={
            yshift=-2mm, 
            xshift=5mm 
        },
        boxed title style={
            colframe=green!75!black, 
            colback=white, 
            sharp corners 
        }
    }
}

\newtcolorbox[auto counter, number within=section]{example}[2][]{%
    examplebox,
    title=Prompt   ~\thetcbcounter~(#2), 
    #1 
}

\usepackage{minitoc}

\title{{\color{title}{PosterReward}}: Unlocking Accurate Evaluation for High-Quality Graphic Design Generation}

\author{%
\parbox{\textwidth}{\centering
Jianyu Lai$^{1,2*}$, Sixiang Chen$^{1,2*}$, Jialin Gao$^{2*}$, Hengyu Shi$^{2}$, Zhongying Liu$^{2}$,\\[2mm]
Fuxiang Zhai$^{1}$, Junfeng Luo$^{2}$, Xiaoming Wei$^{2}$, Lujia Wang$^{1}$, Lei Zhu$^{1,3\dagger}$
}}

\affiliation{%
\parbox{\textwidth}{\centering\small
$^1$The Hong Kong University of Science and Technology (Guangzhou), \quad
$^2$Meituan \\
\quad
$^3$The Hong Kong University of Science and Technology
}}

\contribution[*]{Equal contribution}
\contribution[\dagger]{Corresponding Author}

\abstract{
Recent advancements in the text-rendering capabilities of image generation models have made the end-to-end creation of graphic design content, such as posters, increasingly feasible. However, existing reward models fall short of accurately assessing design quality, as they primarily focus on global image aesthetics while overlooking the critical dimensions of typography and layout. Furthermore, the scarcity of domain-specific preference data remains a significant bottleneck, limiting the further development of graphic design evaluation and generation.

To bridge this gap, we design an automated pipeline to construct a high-quality dataset of 70k poster preferences by leveraging the consensus of multiple Multi-modal Large Language Models (MLLMs) to simulate human-like judgment. Based on this dataset, we propose \textbf{PosterReward}, a reward model specifically designed for high-precision poster assessment through a cascaded, multi-stage training strategy. We also provide multiple variants of the model to cater to different application scenarios. Finally, we introduce \textbf{PosterRewardBench} and \textbf{PosterBench} to evaluate the performance of existing reward models in poster assessment and the generation capabilities of current text-to-image models in poster creation, respectively.
}
\date{\today}
\checkdata[Project Page]{\url{https://alexlai2860.github.io/PosterReward/}}

\begin{document}
\maketitle
\vspace{-8mm}
\begin{figure}[hbt!]
    \centering
    \includegraphics[width=\textwidth]{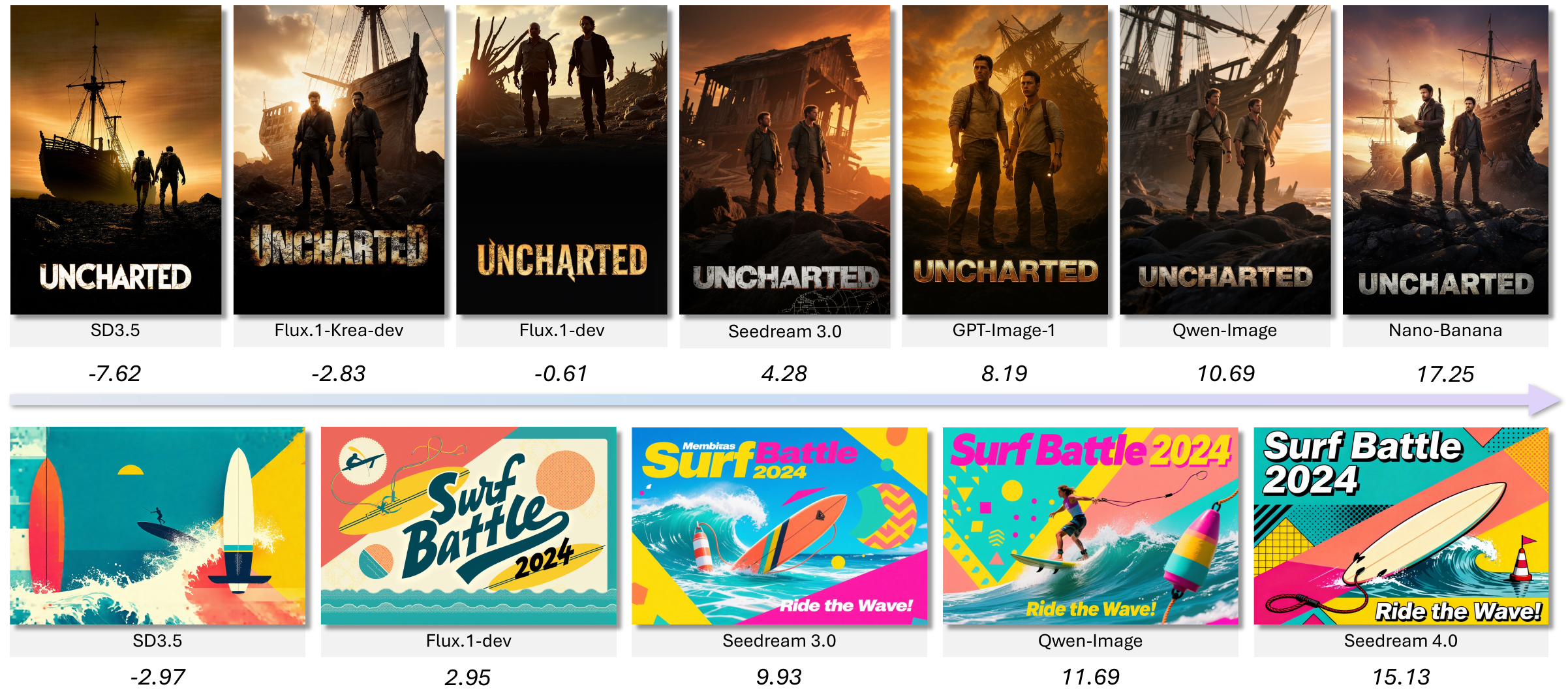}
    \vspace{-0.7cm} 
    \caption{
        \textbf{PosterReward.} 
        PosterReward is a reward model for poster generation tasks. It evaluates posters from multiple dimensions and outputs scores, achieving an accurate assessment of graphic design quality.
    }
    \label{fig:teaser}
\end{figure}
\section{Introduction}
Poster creation is a quintessential application of graphic design, widely utilized in diverse fields such as film, commerce, and public service announcements. An effective poster requires not only a high-quality image but also accurate, aesthetically pleasing text and a well-composed layout. This multifaceted nature presents a significant challenge for image generation models. Merely producing a vivid image is insufficient; the ability to render text accurately and a sophisticated understanding of layout and composition demand more comprehensive capabilities from the model.

Recently, the text rendering capabilities of advanced text-to-image models, such as Flux\cite{flux2024}, PosterCraft\cite{chen2025postercraft}, Seedream\cite{gao2025seedream3,seedream2025seedream4}, Qwen-Image\cite{qwenimage} and HunyuanImage 3.0\cite{cao2025hunyuanimage}, have seen substantial improvements, paving the way for high-quality poster generation. However, existing evaluation models, like HPSv3 \cite{ma2025hpsv3} and UnifiedReward\cite{wang2025unifiedreward}, remain focused on assessing general image preferences and have not been optimized for graphic design content. One primary reason for this is the scarcity of preference datasets tailored to graphic design. For instance, in the largest available dataset, HPDv3\cite{ma2025hpsv3}, the content is predominantly composed of characters (29.4\%), architecture (18.9\%), and art (18.2\%), with design accounting for only 9.9\%. This absence of a dedicated poster evaluation model also impedes the potential for further targeted optimization of generative models through reinforcement learning post-training. Methods such as Flow-GRPO\cite{liu2025flowgrpo}, Pref-GRPO\cite{wang2025prefgrpo}, and Diffusion-NFT\cite{zheng2025diffusionnft} all depend on a reliable reward model to provide the necessary optimization signals.

To address these issues, we first design a reliable preference data collection framework and introduce the Poster-Preference-70K dataset. Unlike traditional methods\cite{kirstain2023pick,xu2023imagereward,ma2025hpsv3} that rely on manual human annotation, our framework is fully automated, leveraging existing reward models, multimodal models, and Multimodal Large Language Models (MLLMs) to generate annotations, using AI-generated preferences as a proxy for human judgments. Our experiments demonstrate that this framework can accurately annotate image preferences while maintaining a high degree of consistency with human evaluators.

The inherent complexity of posters necessitates evaluation across multiple dimensions. Accordingly, we define a five-dimensional evaluation system: Foundational Visual Quality, AI Artifacts, Textual Accuracy, Prompt Fidelity, and Aesthetic Value. While these dimensions are analytically distinct, they are intricately coupled in the final preference judgment and cannot be reduced to a simple weighted average. This interdependence requires a superior poster reward model to be capable of analyzing these dimensions holistically, weighing their respective trade-offs to produce a well-reasoned overall score.

To imbue our reward model with this analytical capability, we develop distinct training strategies for different model architectures. For our generative reward model, PosterReward-Pairwise, we adapt the approach from RewardDance \cite{wu2025rewarddance}, training it to generate a binary judgment (Yes/No) followed by a Chain-of-Thought (CoT) that details the analysis across the five dimensions and the subsequent decision process. During inference, the preference scores for two images can be indirectly computed from the logits of the judgment token. To overcome the limitations of existing discriminative reward models, which often lack analytical depth and test-time scaling, we design the two-stage PosterReward model. This model features an analysis module that assesses an image along the five dimensions and explicitly provides this textual analysis as auxiliary information to a scoring module, thereby enhancing scoring accuracy. Furthermore, we design a multi-stage, cascaded training pipeline that unifies the training of both PosterReward-Pairwise and PosterReward.

Finally, to comprehensively benchmark the poster generation capabilities of existing models, we propose PosterBench, which evaluates the outputs of various models across multiple poster categories using our PosterReward model. Concurrently, to assess the accuracy of existing reward models and MLLMs on the task of poster evaluation, we construct PosterRewardBench, which leverages human-annotated preferences for benchmarking. Experimental results confirm that PosterReward delivers highly accurate preference judgments for posters, outperforming all current reward models and MLLMs.

Our main contributions are as follows:
\begin{itemize}
    \item We introduce the first fully automated preference data collection framework capable of constructing a reliable AI-driven poster preference dataset from unlabeled raw data. This framework significantly reduces the cost of human annotation and offers a blueprint for data collection in other specialized domains.
    \item We design a cascaded, multi-stage reward model post-training pipeline. Centered on multi-dimensional image analysis, it integrates the training of both generative and discriminative reward models, achieving synergistic optimization between these tasks.
    \item We propose \textbf{PosterReward}, the first reward model specifically dedicated to evaluating the quality of generated posters. We also provide several variants to accommodate different use cases.
    \item We introduce two new benchmarks: \textbf{PosterBench}, to measure the performance of text-to-image models on the poster generation task, and \textbf{PosterRewardBench}, to evaluate the accuracy of existing models on poster assessment.
\end{itemize}
\section{Related Works}

\subsection{Pointwise Reward Modeling}
Reward models can be broadly categorized into pointwise and pairwise approaches, with pointwise models being the dominant paradigm. A pointwise reward model takes an image and its prompt as input and outputs a scalar score assessing the image quality. These models can be further divided into two structural subtypes.

\noindent{\textbf{Regressive Reward Modeling.}} This category, often referred to as discriminative reward modeling, directly outputs a scalar score. Early methods were typically built on CLIP\cite{radford2021clip} or its variants, computing image–text quality via cosine similarity, as exemplified by ClipScore\cite{hessel2021clipscore}, PickScore\cite{kirstain2023pick}, HPSv2\cite{hpsv2}, and ImageReward\cite{xu2023imagereward}. InstructGPT\cite{instructgpt} later introduced a simple regression head that replaces the final language-model layer, shifting the output from logits to a scalar value for RLHF preference alignment. This design was quickly adopted by multimodal reward models, including IXC-2.5-Reward\cite{ixc2.5}, Skywork-vl Reward\cite{wang2025skyworkvlreward}, and BaseReward\cite{zhang2025basereward}, and was extended to visual generation assessment by VideoScore\cite{he2024videoscore}. The recent HPSv3\cite{ma2025hpsv3} also employs this architecture and achieves state-of-the-art results across multiple benchmarks. All these models are trained on large-scale, human-annotated preference pairs using the Bradley–Terry loss\cite{bradley1952rank}.

\noindent{\textbf{Generative Reward Modeling.}} This paradigm reframes reward prediction as a token-generation task, naturally aligning with VLM pre-training. It has been widely used in visual-generation reward models as well as Image Quality Assessment (IQA) and Image Aesthetics Assessment (IAA). One line of work fine-tunes a VLM to output the score as a text string that is later parsed into a numerical value, as done in UnifiedReward\cite{wang2025unifiedreward,wang2025unifiedrewardthink}, EditScore\cite{luo2025editscore}, Q-Insight\cite{li2025qinsight}, VisualQuality-R1\cite{wu2025visualquality}, and Next Token Is Enough\cite{li2025next}. A more flexible alternative derives the reward from the model’s logits over predefined categorical choices: the VLM evaluates the image using these categories, and the resulting probability distribution is mapped to a final score. Representative examples include RewardDance\cite{wu2025rewarddance}, VQA-score\cite{lin2024vqascore}, Q-Align\cite{wu2023qalign}, DeQA-Score\cite{deqa}, SnowMaster\cite{lai2025snowmaster}, and Artimuse\cite{cao2025artimuse}.

\subsection{Pairwise Reward Modeling}

A pairwise reward model takes two images as input and outputs either individual scores or their relative ranking. This paradigm was first introduced to visual generation by UnifiedReward\cite{wang2025unifiedreward}. The series has since evolved: UnifiedReward 2.0 evaluates multiple dimensions of the two images separately, while UnifiedReward-think\cite{wang2025unifiedrewardthink} improves performance through post-training techniques such as Chain-of-Thought, rejection sampling, and reinforcement learning. More recently, RewardDance\cite{wu2025rewarddance} explores the scaling potential of pairwise RMs. The paradigm has also expanded to new domains, with OneReward\cite{gong2025onereward} adapting it to image editing by using a single model to support multiple editing tasks.

\subsection{Preference Datasets}

The performance of reward models is closely tied to the scale and quality of the preference datasets used for training. In visual generation, several key datasets have been developed to capture human preferences. Early human-annotated datasets such as HPDv1\cite{wu2023hpdv1}, ImageRewardDB\cite{xu2023imagereward}, and Pick-a-Pic\cite{kirstain2023pick} focused on pairwise comparisons between images generated from the same prompt, primarily evaluating outputs from existing models. HPDv2\cite{hpsv2} expanded the data diversity by incorporating images from a broader set of generative models. HPDv3\cite{ma2025hpsv3} further scales this effort, providing a wide-spectrum text-to-image preference dataset with 1.08M text–image pairs and 1.17M annotated comparisons, combining outputs from many state-of-the-art models with high-quality real images. Traditional human annotation incurs substantial time and labor costs. With the growing capabilities of advanced models, some works\cite{wu2024visionprefer} have explored MLLM-based automated evaluation, but methods like multi-image ranking demand strong model capabilities and can introduce positional bias, limiting their suitability for today’s complex, high-quality visual content. To address these challenges, recent efforts such as Skywork-Reward-V2\cite{liu2025skywork} explore more robust human–AI synergistic preference evaluation with LLM assistance.

\subsection{Graphic Design Evaluation}
Automatic graphic design evaluation has transitioned from handcrafted heuristic metrics \cite{o2014learning, kong2022aesthetics++} to foundation-model-driven approaches. Recent studies investigate the capability of Large Multimodal Models (LMMs) in assessing specific design principles \cite{haraguchi2024can} and aligning layout generation with human aesthetic preferences \cite{patnaik2025aesthetiq, goyal2025design}. However, these methodologies remain predominantly layout-centric, focusing on the geometric arrangement of elements while largely overlooking the intricate quality of textual rendering and global aesthetics.

While recent agentic frameworks \cite{nag2025agentic} attempt multi-faceted assessment, they often rely on decoupled agents that provide fragmented feedback, struggling to capture the holistic synergy between structure, typography, and visual beauty. In contrast, our PosterReward introduces a unified evaluation paradigm. By integrating structural layout, text-rendering accuracy, and aesthetic expression into a single balanced score, PosterReward provides a comprehensive reward signal. This multidimensional fusion allows for a more rigorous optimization of overall design quality than existing layout-only or fragmented evaluation schemes.
\vspace{-0.1cm}
\section{Dataset Pipeline}

\begin{figure*}[t!]
    \centering
    \includegraphics[page=1, width=\textwidth]{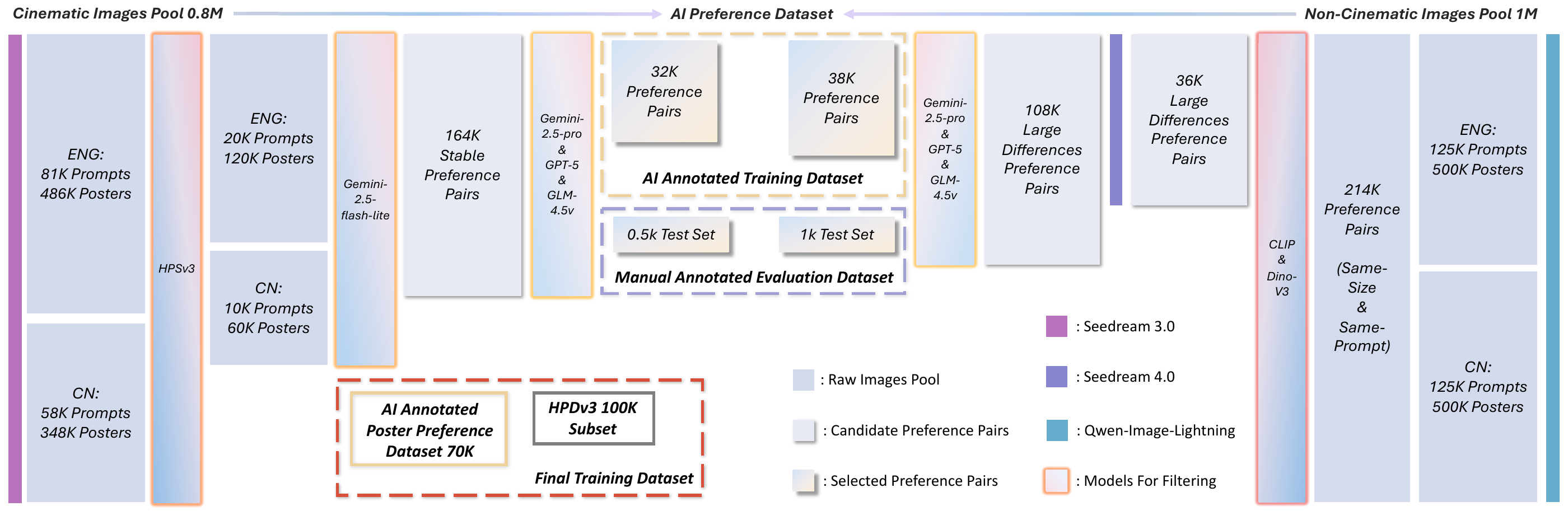}
    \caption{Schematic diagram of AI preference data collection. The raw data was generated using Seedream 3.0, Seedream 4.0, and Qwen-Image-Lightning. The models used included four open-source models: CLIP, DINOv3, HPSv3, and GLM-4.5v, and three closed-source models: Gemini-2.5-Flash-Lite, Gemini-2.5-Pro, and GPT-5.}
    \label{fig_datapipeline}
\end{figure*}

\begin{figure}[t!]
    \includegraphics[width=\textwidth]{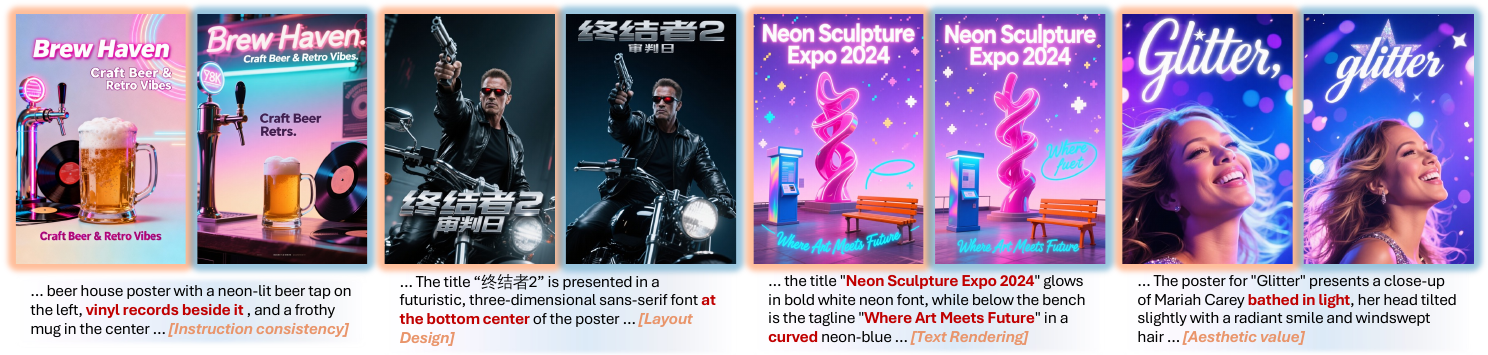}
    \caption{A schematic diagram of AI preference data samples. In each group of images, the left side represents chosen samples, and the right side represents rejected samples. The orange box at the end of the prompt section below indicates the dimensions used to construct the preference pairs.}
    \label{fig_samples}
\end{figure}

Existing human preference datasets in the visual generation domain, such as Pick-a-Pic\cite{kirstain2023pick}, ImageReward\cite{xu2023imagereward}, and HPDv3\cite{ma2025hpsv3}, predominantly feature preference data for a broad spectrum of generated images, including paintings and photographs. However, graphic design content is significantly underrepresented. To address this gap, we constructe the \textbf{Poster-Preference-70K} dataset, which focuses specifically on graphic design preferences.

\subsection{AI Preference Data Collection Pipeline}
Traditional human preference data collection is slow, expensive, and requires extensive annotator training to ensure quality.  With the rapid advancement of Multimodal Large Language Models (MLLMs), the ``MLLM-as-a-judge'' paradigm has been widely adopted across various domains. This leads us to investigate the feasibility of using MLLMs to replace human annotators in building a reliable AI-judged preference dataset.

We propose an automated framework for constructing preference data. As illustrated in Figure~\ref{fig_datapipeline}, This framework leverages several advanced multimodal models and reward models to generate high-quality AI-judged preferences from unlabeled raw data. Specifically, for two distinct sample pools---cinematic and non-cinematic---we design separate pipelines for initial filtering and preference pairing. Both pipelines culminate in a unified multi-model selection framework for final preference pair validation. The overall pipeline architecture is illustrated in Figure~\ref{fig_datapipeline}. For preference samples, please refer to Figure~\ref{fig_samples}. The left part of the figure depicts the pipeline for cinematic images. The initial data pool contains 0.8 million movie poster images generated by Seedream 3.0\cite{gao2025seedream3} from 81K English and 58K Chinese prompts, with each prompt yielding six images. For non-cinematic content, which includes various commercial and public service graphic designs, the data pool comprises images generated by Qwen-Image-Lightning\cite{qwenimage} from 125k English and 125k Chinese prompts respectively, with four images per prompt. To manage these large-scale data pools efficiently, we adopt a cascaded-model approach to reduce construction costs while ensuring the reliability of the final dataset.

\noindent{\textbf{Cinematic Data.}} 
For the cinematic poster data, we observe that it contains a large proportion of portraits and exhibits distinct aesthetic styles, which aligns well with the training data of HPSv3\cite{ma2025hpsv3}. Consequently, we chose HPSv3 to score each image in the first filtering stage. To ensure the reliability of these preferences, we employ Kendall's Coefficient of Concordance (Kendall's $W$) \cite{kendall1939problem} to quantify the consistency across the six ranking iterations, treating each round as a pseudo-rater. We select the top 20k English and 10k Chinese prompt groups with the highest $W$ values, representing the most stable model consensus. From these 30k groups, 450k potential preference pairs could be formed. As directly processing this volume would be resource-intensive, we introduce an intermediate step using a lightweight, closed-source model to perform multiple rankings. In our implementation, we perform six ranking iterations and selected pairs where the relative order was consistent in at least five of them. This process yields 164k candidate pairs for the final multi-model filtering stage.

\noindent{\textbf{Non-Cinematic Data.}} 
For the non-cinematic images, which generated by Qwen-Image-Lightning could have inconsistent dimensions, we first filter for pairs with identical dimensions, resulting in 214k candidate pairs. Since these images often exhibit low variance, we implement a subsequent filtering step to ensure significant dissimilarity within each selected pair, using both CLIP\cite{radford2021clip} and DINOv3\cite{simeoni2025dinov3}. CLIP assesses semantic similarity, while DINOv3 focuses on structural similarity. We select the union of the top 15k most dissimilar pairs according to DINOv3 and the top 25k according to CLIP, yielding 36k high-variance pairs. As this quantity was insufficient, we expand the dataset with 36k high-quality poster images generated by Seedream 4.0 \cite{seedream2025seedream4} from which we form 108k new candidate pairs.

\noindent{\textbf{Multi-Model Filtering.}} 
For the final preference pair validation, to mitigate single-model bias, we employ a panel of three state-of-the-art MLLMs: Gemini-2.5-Pro\cite{comanici2025gemini}, GPT-5, and GLM-4.5v\cite{vteam2025glm45vglm41vthinkingversatilemultimodal}. These models jointly select reliable preference pairs via pairwise comparison. During this process, we observe that current MLLMs exhibit a significant positional bias, where they tend to favor the first image presented over the second (see Table~\ref{tab:pairwise_comparison_prb_compact}). To counteract this bias, we prompt each model with every pair twice, swapping the image order for the second evaluation.

\section{Method}

\begin{figure*}[!t]
    \centering
    \includegraphics[width=\linewidth]{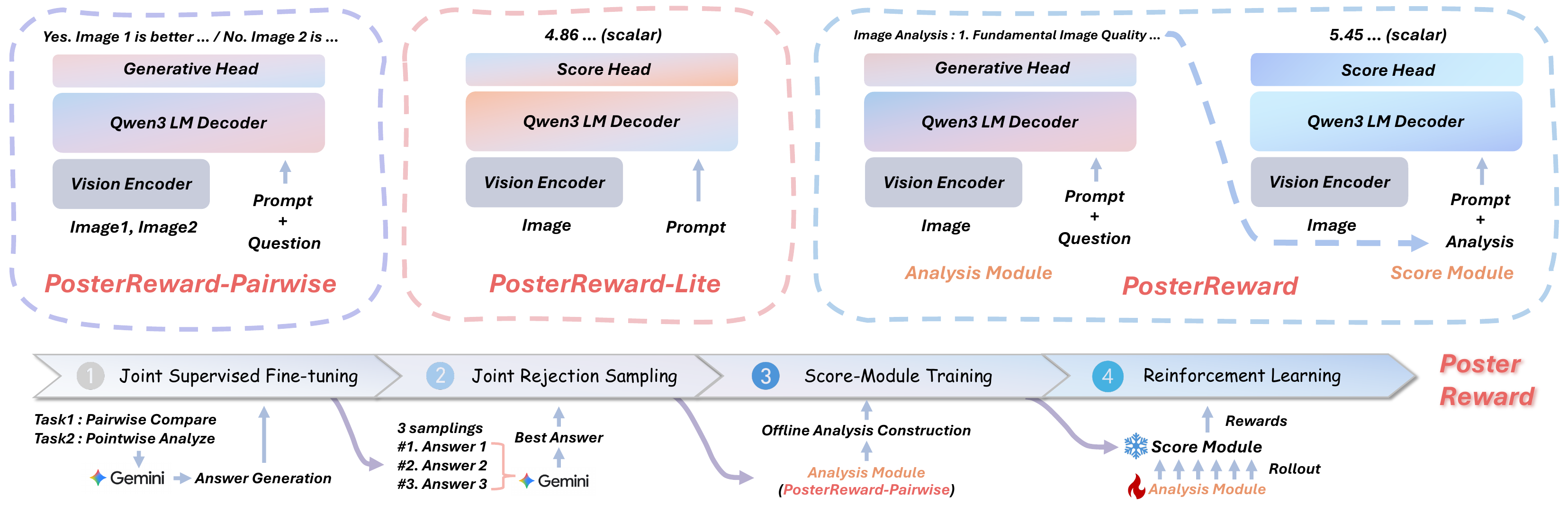}
    \caption{PosterReward training pipeline and model structure diagram. The top shows three reward models with different structures, and the bottom shows the training pipeline. Our training pipeline consists of four cascaded stages: Joint Supervised Fine-Tuning, Joint Rejection Sampling, Score-Module Training, and Reinforcement Learning.}
    \label{fig_framework}
\end{figure*}

\subsection{Model Architecture}

Pointwise models offer flexibility and have been widely applied in reinforcement learning for image generation. Generative pointwise models typically follow two main approaches: regressing scores directly from labeled data \cite{wang2025unifiedreward} or indirectly calculating scores from logits \cite{wu2025rewarddance}. The former is susceptible to annotator bias and incurs high labeling costs. The latter, which relies on human-defined labels, often fails to align impartially with real-world image distributions, leading to annotation bias \cite{deqa}. We contend that the optimal solution is a discriminative pointwise reward model, a methodology whose reliability has been validated in the LLM and VLM domains. However, existing discriminative models often lack interpretability and cannot improve performance via test-time scaling, as they rely solely on their intrinsic understanding.

By inheriting the complete architecture of VLMs, pairwise reward models can leverage techniques such as Chain-of-Thought (CoT) and reinforcement fine-tuning to enhance performance \cite{wang2025unifiedrewardthink, wu2025rewarddance}. They also exhibit greater interpretability and potential for scaling \cite{wu2025rewarddance}, making them well-suited for tasks like preference data filtering. Nevertheless, their application in RL-based fine-tuning for image generation is limited by the high computational cost of performing pairwise comparisons across numerous samples.

To address the distinct advantages and limitations of each model type, we design a cascaded training pipeline focused on image analysis. This pipeline integrates a discriminative model, PosterReward, and a pairwise model, PosterReward-Pairwise.

\noindent{\textbf{PosterReward.}} We design a novel two-stage discriminative reward model that incorporates image analysis and test-time scaling capabilities. An image and its corresponding prompt are first fed into a first-stage analysis module, fine-tuned from Qwen3-VL-8B\cite{yang2025qwen3}, which generates a multi-dimensional analysis. Subsequently, the image, the analysis text, and the prompt are passed to a second-stage scoring module.  For this module, we follow the suggestion from BaseReward~\cite{zhang2025basereward} by replacing the final layer of Qwen3-VL-8B with a two-layer MLP, connected by a SiLU activation, to output a final scalar score.

\noindent{\textbf{PosterReward-Lite.}} A simplified variant of PosterReward, which omits the analysis module for inference scenarios where computational speed is critical.

\noindent{\textbf{PosterReward-Pairwise.}} This generative reward model is fine-tuned entirely from Qwen3-VL-8B. Following the methodology of RewardDance \cite{wu2025rewarddance}, we train the model to first make a preference judgment and then output its CoT reasoning. This approach prevents the corruption of the logits distribution that can occur when the judgment is produced after the reasoning process \cite{li2025uniworld}. During inference, the full CoT can be omitted to accelerate processing.
\subsection{Training Pipeline}

We propose a unified training pipeline composed of four cascaded stages: Joint Supervised Fine-Tuning, Joint Rejection Sampling Fine-Tuning, Scoring Module Training, and Reinforcement Learning Fine-Tuning.

\noindent{\textbf{Joint Supervised Fine-Tuning.}} We use Gemini-2.5-Pro\cite{comanici2025gemini} to annotate our raw data for two joint tasks:
\begin{enumerate}[label=(\arabic*)]
    \item \textit{Single-Image Analysis}: The model is prompted to provide a comprehensive and objective evaluation of an image from multiple dimensions.
    \item \textit{Paired-Image Comparison}: The model is required to first state its preference between two images based on the same dimensions, and then provide a detailed CoT for its decision. Notably, we randomly swapped the positions of the ``chosen'' and ``rejected'' images to ensure a balanced distribution of ``Yes'' and ``No'' responses. This strategy can balance the number of the two types of responses and reduce the inherent position bias of the model.
\end{enumerate}
We have collected 246k single-image analysis samples and 160k paired-preference samples to fine-tune Qwen3-VL-8B. We posit that these two tasks are intrinsically linked: learning to judge image quality enhances the model's analytical capabilities, while a comprehensive analysis enables more accurate preference judgments.

\noindent{\textbf{Joint Rejection Sampling Fine-Tuning.}} From the SFT-tuned model, we sample three distinct responses for each prompt across both tasks. We then use Gemini-2.5-Flash-Lite\cite{comanici2025gemini} to select the highest-quality response for rejection sampling fine-tuning. For cases where the SFT model made an incorrect preference judgment, we replace the chosen response with the original ground-truth response from Gemini-2.5-Pro. This stage further enhances the model's output quality. The resulting model, which excels at both paired comparison and single-image analysis, serves as the final \textbf{PosterReward-Pairwise} model and the initial analysis module for the fourth stage.

\noindent{\textbf{Scoring Module Training.}} In this stage, we utilize the model from the second stage to re-annotate the analysis texts within our preference dataset. Subsequently, the scoring module is trained on preference pairs, where each training instance is formulated as a triplet. Formally, we denote the triplets for the chosen (winning) and rejected (losing) samples as $x_w$ and $x_l$, respectively:
\begin{equation}
x_w = (I_w, P, A_w), \quad x_l = (I_l, P, A_l)
\end{equation}
where $I_w$ and $I_l$ represent the chosen and rejected images, $P$ is the shared prompt, and $A_w, A_l$ denote their corresponding analysis texts. The training process is optimized using the Bradley-Terry loss function:
\begin{equation}
\mathcal{L}_{BT} = -\mathbb{E}_{(x_w, x_l) \sim \mathcal{D}} \Big[ \log \sigma \left( r_\theta(x_w) - r_\theta(x_l) \right) \Big]
\end{equation}

\noindent{\textbf{Reinforcement Learning.}} In the final stage, we fine-tune the analysis module using Group Relative Policy Optimization (GRPO)\cite{shao2024deepseekmath}, with the frozen scoring module acting as the reward function. This process enhances the quality of the generated analysis texts.

First, we define a reward $r_i$ for each sample based on its preference status. For a chosen (winning) sample, the reward is the score from the scoring module; for a rejected (losing) sample, it is the negative of the score:
\begin{equation}
r_i =
\begin{cases}
r(I_w, P, A_w) & \text{if sample } i \text{ is preferred} \\
-r(I_l, P, A_l) & \text{if sample } i \text{ is rejected}
\end{cases}
\end{equation}
This reward is then normalized across the batch to compute the advantage $\hat{A}_i$, which represents the relative quality of the analysis:
\begin{equation}
\hat{A}_{i} = \frac{r_i - \text{mean}(\mathbf{r})}{\text{std}(\mathbf{r})}
\end{equation}
where $\mathbf{r}$ is the vector of rewards for all samples in the batch.

With the advantage defined, we optimize the analysis model $\pi_\theta$ using the GRPO objective. The objective function incorporates a clipped surrogate objective using the probability ratio $\rho_i(\theta)$ to stabilize training, along with a KL divergence penalty to regularize the policy against a reference model $\pi_{\text{ref}}$. The final objective is:
\begin{equation}
\begin{aligned}
\mathcal{L}_{\text{GRPO}}(\theta) = \mathbb{E}\Big[ &\min\big(\rho_i(\theta)\hat{A}_i, \text{clip}(\rho_i(\theta), 1-\delta, 1+\delta)\hat{A}_i\big) \\
& - \beta D_{KL}(\pi_\theta || \pi_{\text{ref}}) \Big]
\end{aligned}
\end{equation}
This GRPO process encourages the analysis module to produce descriptions that align better with the scoring module's criteria, leading to a more robust final PosterReward model.

\noindent{\textbf{Summary of the Pipeline.}} In essence, our pipeline leverages image analysis to synergistically enhance both reward models. Joint training with analytical tasks directly increases the performance of \textbf{PosterReward-Pairwise}, while the explicit integration of the refined analysis module significantly improves the scoring quality of \textbf{PosterReward}.
\vspace{-0.1cm}
\section{Experiments}
\vspace{-0.1cm}

\subsection{PosterRewardBench}

\subsubsection{Experiment Setup}

Experiments for the first three stages are conducted on eight A100 GPUs. In joint Supervised Fine-Tuning (SFT) and joint Rejection Sampling Fine-Tuning (RSFT), we employ full-parameter fine-tuning with a learning rate of $1 \times 10^{-4}$. For the third stage, scoring module training, we utilize LoRA (rank 64) at the same learning rate. The fourth stage, GRPO fine-tuning, also employs LoRA (rank 64) with a learning rate of $1 \times 10^{-6}$, utilizing 8 A100s for training, 8 for rollout, and 4 for reward model deployment. Further training details are provided in the supplementary materials.
\vspace{-0.2cm}
\subsubsection{Benchmark Details}

PosterRewardBench is designed to measure the accuracy of existing MLLMs and reward models in evaluating poster preferences. We provide two benchmarks constructed using different models: \textbf{PosterRewardBench-Basic} and \textbf{PosterRewardBench-Advanced}. The former is generated by Flux, Flux-Krea, and SD3.5-L, exhibiting significant variations in average sample quality. The latter is generated by Seedream3.0\cite{gao2025seedream3}, Seedream4.0\cite{seedream2025seedream4}, and Qwen-Image-Lightning\cite{qwenimage}, featuring higher overall quality and smaller quality differences among samples. All preference pairs were reviewed by four professional annotators, and only those with a consensus from at least three annotators were retained. For more detailed statistics, please refer to the supplementary materials. Furthermore, we test the models' judgment capabilities on general image generation using the HPDv3\cite{ma2025hpsv3} and MMRB2\cite{hu2025multimodal} benchmarks. For the pairwise evaluation of both reward models and MLLMs, we conduct separate assessments with swapped image positions to investigate potential positional bias.

\subsubsection{Evaluation Results}

\begin{table}[!ht]
\centering
    \setlength{\abovecaptionskip}{0.1cm} 
\captionsetup{font=small}
\caption{Performance comparison of \textbf{pointwise reward models} across various benchmarks. All values represent accuracy ($\uparrow$). PRB is an abbreviation for PosterRewardBench.}
\label{tab:pointwise_reward_comparison_compact}
\setlength{\tabcolsep}{4pt} 
{\footnotesize
\renewcommand\arraystretch{1.2}
\begin{tabular*}{0.8\linewidth}{@{\extracolsep{\fill}}l cccc@{}}
\toprule
\textbf{Model} & \textbf{MMRB2} & \textbf{HPDv3} & \textbf{PRB-Basic} & \textbf{PRB-Ad} \\
\midrule
ImageReward\cite{xu2023imagereward}       & 53.0 & 58.6 & 60.7 & 49.3 \\
PickScore\cite{kirstain2023pick}         & 57.6 & 65.6 & 66.7 & 44.1 \\
HPSv2\cite{hpsv2}             & 55.0 & 65.3 & 70.8 & 43.7 \\
UnifiedReward*\cite{wang2025unifiedreward}     & 56.9 & 59.4 & 60.0 & 52.7 \\
HPSv3\cite{ma2025hpsv3}             & 58.5 & 76.9 & 72.9 & 41.2 \\
\midrule
PosterReward-Lite & \textbf{60.5} & \underline{77.1} & \underline{83.9} & \underline{85.0} \\
\textbf{PosterReward} & \underline{59.6} & \textbf{77.8} & \textbf{86.7} & \textbf{86.0} \\
\bottomrule
\end{tabular*}
}
\end{table}

\begin{table}[t!]
\centering
    \setlength{\abovecaptionskip}{0.1cm} 
\captionsetup{font=small}
\caption{Performance comparison of \textbf{pairwise reward models} on PosterRewardBench (PRB). ``Yes'' and ``No'' refer to the accuracy on samples with positive and negative ground truth labels, respectively. All values represent accuracy ($\uparrow$).}
\label{tab:pairwise_comparison_prb_compact}
\setlength{\tabcolsep}{4pt} 
{\footnotesize
\renewcommand\arraystretch{1.2}
\begin{tabular*}{0.8\linewidth}{@{\extracolsep{\fill}}l ccc ccc@{}}
\toprule
& \multicolumn{3}{c}{\textbf{PRB-Basic Acc.} $\uparrow$} & \multicolumn{3}{c}{\textbf{PRB-Ad. Acc.} $\uparrow$} \\
\cmidrule(lr){2-4} \cmidrule(lr){5-7}
\textbf{Model} & \textbf{Yes} & \textbf{No} & \textbf{Avg.} & \textbf{Yes} & \textbf{No} & \textbf{Avg.} \\
\midrule
UnifiedReward-think\cite{wang2025unifiedrewardthink}   & 75.1 & 61.5 & 68.3 & 52.6 & 48.5 & 50.6 \\
Qwen3-VL-Plus\cite{yang2025qwen3}              & 89.9 & 39.2 & 64.5 & 98.7 & 14.2 & 56.4 \\
Gemini-2.5-Flash\cite{comanici2025gemini}      & 94.7 & 33.3 & 64.0 & 95.2 & 28.8 & 62.0 \\
Gemini-2.5-Pro\cite{comanici2025gemini}        & 75.6 & 83.1 & 79.3 & 81.8 & 68.6 & 75.2 \\
GPT-5                & 90.4 & 80.5 & \textbf{85.4} & 89.8 & 75.9 & \underline{82.9} \\
\midrule
\textbf{PosterReward-Pairwise} & 82.0 & 84.0 & \underline{83.0} & 84.1 & 83.6 & \textbf{83.8} \\
\bottomrule
\end{tabular*}
}
\end{table}

Table~\ref{tab:pointwise_reward_comparison_compact} presents the comprehensive evaluation results. For UnifiedReward, we utilize the Qwen-7B version. Given that UnifiedReward produces a high frequency of ties in pointwise mode, we calculate its accuracy by treating each tie as half a correct prediction ($Accuracy = N_{correct} + 0.5 \times N_{ties}$), denoted as $UnifiedReward^*$. The results demonstrate that the PosterReward series consistently achieves superior accuracy across various datasets. On PosterReward-Advanced, PosterReward attains an accuracy of 86.5\%, which significantly surpasses existing baselines that mostly range between 40\% and 53\%. This indicates that previous reward models struggle to accurately evaluate high-quality, complex generated content. Moreover, PosterRewardBench-Basic constitutes out-of-distribution (OOD) poster data for PosterReward, yet it still achieves a leading accuracy of 86.7\%, showcasing its strong robustness. Finally, on general text-to-image benchmarks including HPDv3 and the T2I sub-benchmark of MMRB2, PosterReward and its Lite version secure the highest accuracy among all pointwise models, confirming their excellent generalization capabilities.

In the pairwise model evaluation illustrated in Table~\ref{tab:pairwise_comparison_prb_compact}, the jointly trained PosterReward-Pairwise model also exhibits robust performance, achieving the highest scores on PosterRewardBench-Advanced, and is second only to GPT-5 on PosterRewardBench-Basic. More importantly, PosterReward-Pairwise, which has been trained on a balanced dataset, demonstrates minimal positional bias in tests with varied image positions. In contrast, existing MLLMs and UnifiedReward exhibit a certain degree of positional bias.

\subsubsection{Ablation Studies}

\begin{figure}[ht]
    \centering
    \includegraphics[page=1, width=0.9\textwidth]{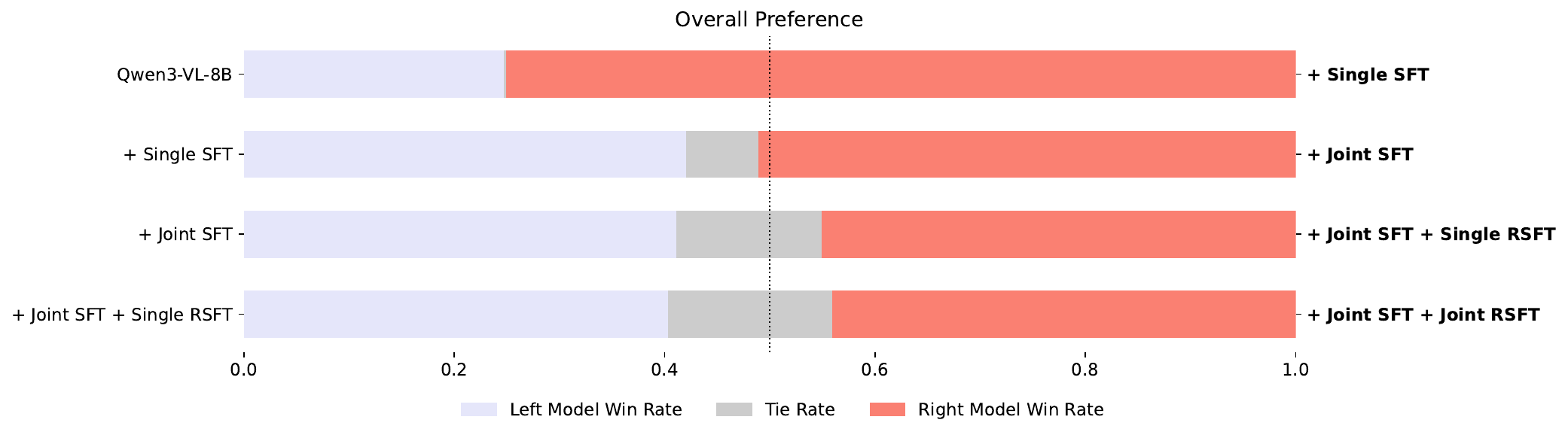}
    \caption{Using the MLLM-as-a-judge method, this figure illustrates our preference analysis for the analysis module via Gemini-3-flash. The results show that the SFT model performs significantly better than the base model, with both joint SFT and joint RSFT contributing to performance gains. We averaged the results of two separate annotations with swapped text positions to eliminate position bias.}
    \label{fig_gemini3flash_preference}
\end{figure}

We first conduct ablation studies on the initial stages of our pipeline to validate the effectiveness of our training strategy. To provide a more human-aligned evaluation of the analysis module, we move beyond traditional reference-based metrics and employ an MLLM-as-a-judge approach. Specifically, we utilize Gemini-3-flash to conduct a pairwise preference analysis across different training stages. As illustrated in Figure~\ref{fig_gemini3flash_preference}, the fine-tuned models significantly outperform the base Qwen3-VL-8B model. Furthermore, the results demonstrate that both joint SFT and joint RSFT contribute to consistent performance gains, validating the synergy between tasks. To ensure evaluation fairness, we averaged results from two separate annotations with swapped text positions to eliminate position bias. The improvement in the pairwise preference judgment task was also directly assessed via preference accuracy, as presented in Table~\ref{tab:ablation_accuracy_pairwise}.

Furthermore, we conduct ablation studies on the PosterReward training pipeline. The ablation results in Table~\ref{tab:ablation_posterreward_focused} show that both the image analysis module and GRPO optimization contribute to consistent performance gains on the HPDv3 and PRB benchmarks. Notably, the accuracy improvements are more pronounced on the poster-related benchmarks (PRB-Basic and PRB-Advanced) compared to the general text-to-image dataset.

\begin{table}[t]
\centering
    \setlength{\abovecaptionskip}{0.1cm} 
    \setlength{\belowcaptionskip}{-0cm}
\captionsetup{font=small}
\caption{Ablation Study on PosterReward-pairwise Model. ``Yes'' and ``No'' refer to the ground truth of the response.}
\label{tab:ablation_accuracy_pairwise}
\setlength{\tabcolsep}{4pt} 
{\footnotesize 
\renewcommand\arraystretch{1.1}
\begin{tabular*}{0.8\linewidth}{@{\extracolsep{\fill}}l ccc ccc@{}}
\toprule
& \multicolumn{3}{c}{\textbf{Advanced Acc.} $\uparrow$} & \multicolumn{3}{c}{\textbf{Basic Acc.} $\uparrow$} \\
\cmidrule(lr){2-4} \cmidrule(lr){5-7}
\textbf{Method} & \textbf{Yes} & \textbf{No} & \textbf{Avg.} & \textbf{Yes} & \textbf{No} & \textbf{Avg.} \\
\midrule
SFT (Single)      & 81.77 & 82.09 & 81.93 & 80.85 & 82.59 & 81.72 \\
SFT (Joint)       & 82.09 & 83.32 & 82.71 & 80.08 & 83.75 & 81.92 \\
+ RSFT (Single)   & 82.67 & 83.24 & 82.96 & 80.66 & 83.56 & 82.11 \\
+ RSFT (Joint)    & \textbf{84.06} & \textbf{83.57} & \textbf{83.82} & \textbf{82.01} & \textbf{83.95} & \textbf{82.98} \\
\bottomrule
\end{tabular*}
}
\end{table}

\begin{table}[t]
\centering
    \setlength{\abovecaptionskip}{0.1cm} 
\captionsetup{font=small}
\caption{Ablation study of PosterReward on key benchmarks, showing the cumulative impact of each component. All values represent accuracy ($\uparrow$).}
\label{tab:ablation_posterreward_focused}
\setlength{\tabcolsep}{10pt} 
{\footnotesize
\renewcommand\arraystretch{1.2}
\begin{tabular*}{0.8\linewidth}{@{\extracolsep{\fill}}l ccc@{}}
\toprule
\textbf{Model / Component} & \textbf{HPDv3} & \textbf{PRB-Basic} & \textbf{PRB-Ad} \\
\midrule
\textit{PosterReward-Lite} & 77.1 & 83.9 & 85.0\\
+ Analysis                   & 77.5 & 85.7 & 85.8\\
\makecell[l]{\textbf{+ Analysis + GRPO} \textbf{ (\textit{PosterReward})}} & \textbf{77.8} & \textbf{86.7} & \textbf{86.0} \\
\bottomrule
\end{tabular*}
}
\end{table}

\subsubsection{Visual Comparison}

\begin{figure*}[!t]
    \centering
    \setlength{\abovecaptionskip}{0.1cm} 
    \includegraphics[page=1, width=\textwidth]{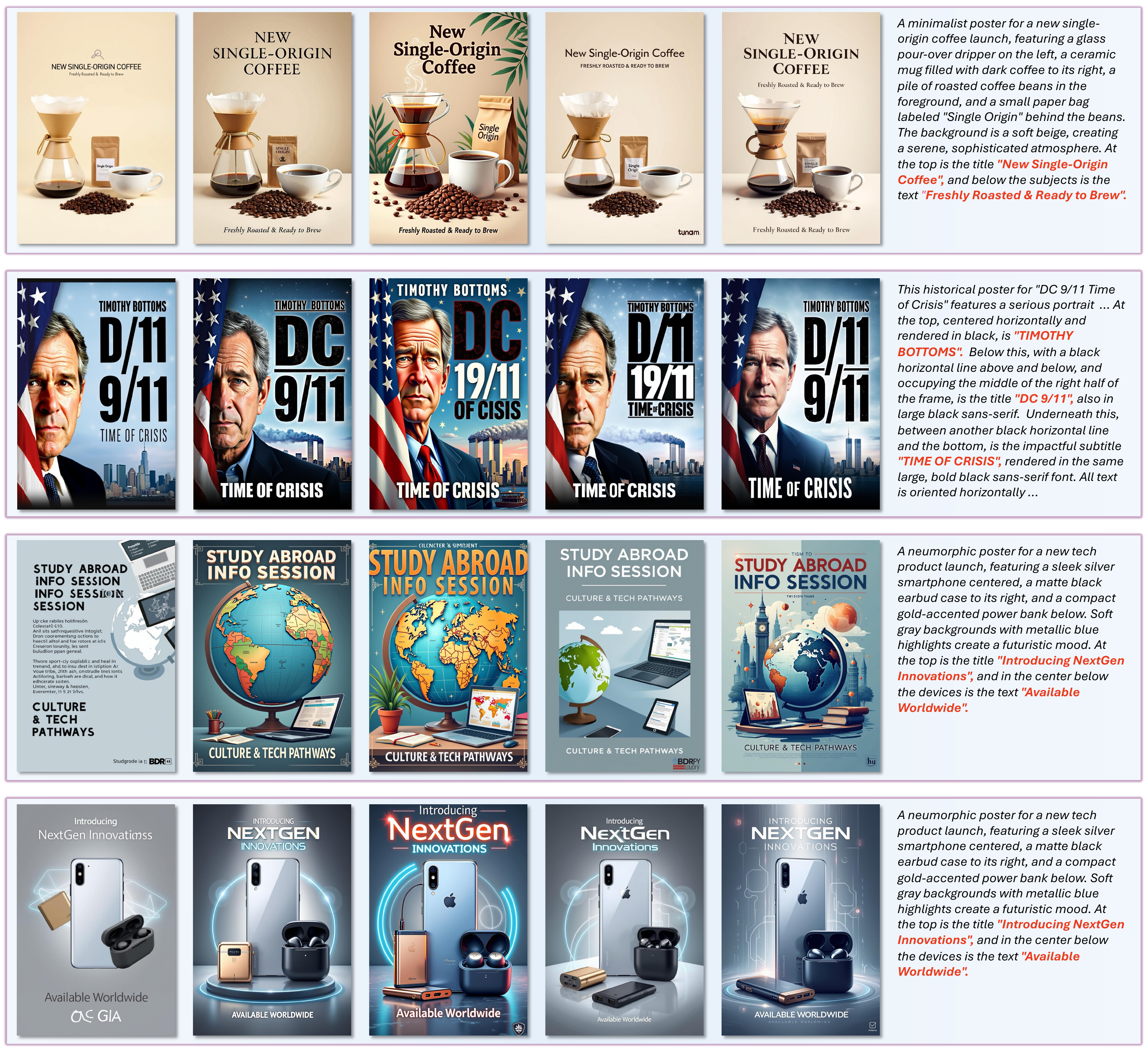}
    \caption{\textbf{Visual comparison of Flux.1-dev fine-tuned with various reward models.} From left to right, the columns display the outputs of the original Flux.1-dev, followed by models fine-tuned using \textbf{PosterReward}, \textbf{HPSv3}, \textbf{UnifiedReward} and \textbf{PickScore}. The corresponding prompts are enclosed in the purple boxes on the right.}
    \label{flux_comp}
\end{figure*}

\begin{figure*}[t!]
    \centering
    \setlength{\abovecaptionskip}{0cm} 
    \vspace{-1cm}
    \includegraphics[page=1, width=\textwidth]{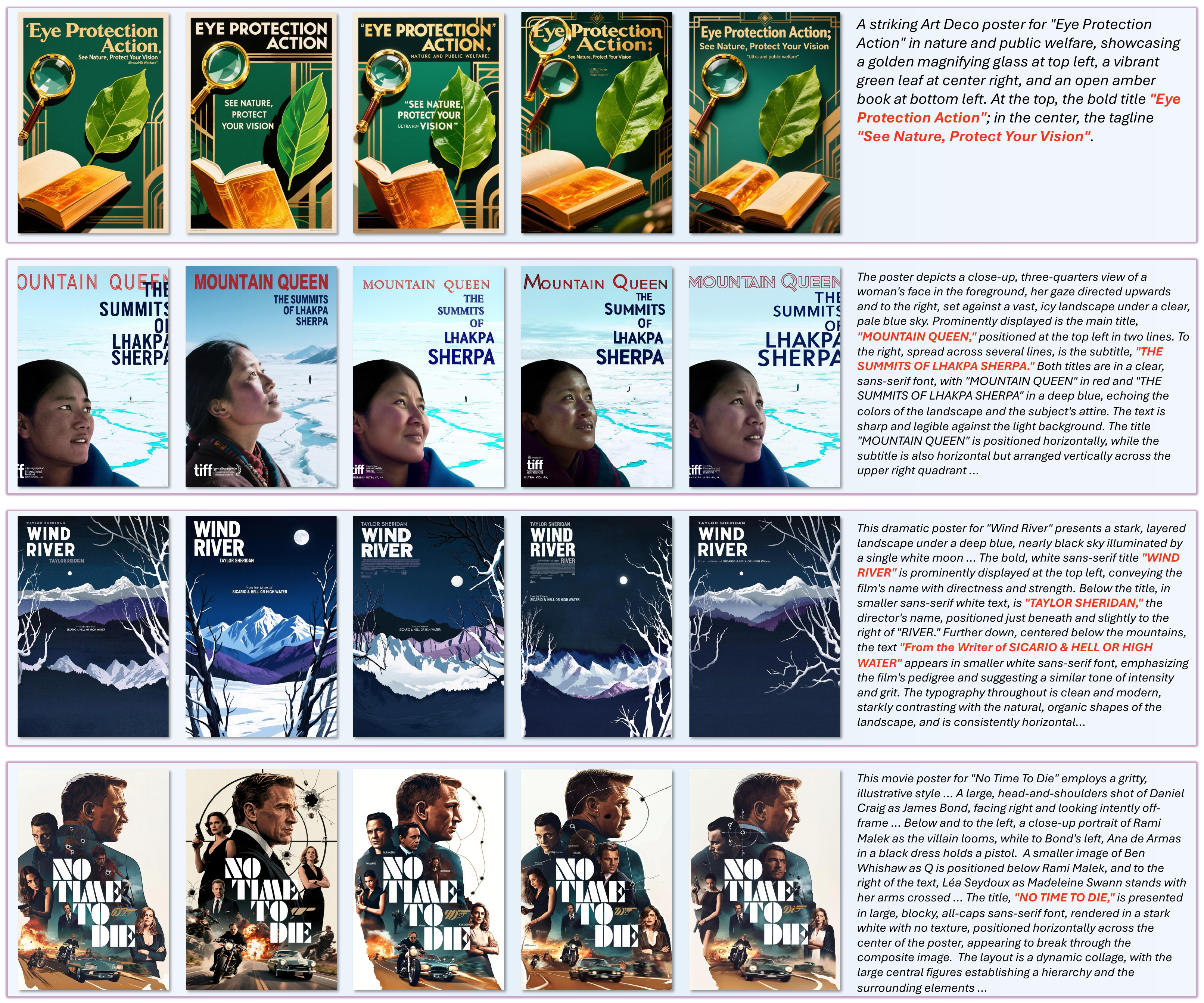}
    \caption{\textbf{Visual comparison of Qwen-Image fine-tuned with various reward models.} Consistent with the previous figure, the columns represent the original Qwen-Image, PosterReward, HPSv3, UnifiedReward and Pickscore respectively. The corresponding prompts are enclosed in the purple boxes on the right.}
    \label{qwen_comp}
\end{figure*}

We evaluate the effectiveness of PosterReward as a reward model for post-training image generation and provide visual comparisons against existing SOTA baselines. Specifically, we fine-tune Flux.1-dev and Qwen-Image utilizing Flow-GRPO\cite{liu2025flowgrpo}; the visual results are presented in figure \ref{flux_comp} and figure \ref{qwen_comp}, respectively. Furthermore, the performance of the SD3.5-M model fine-tuned via DiffusionNFT\cite{zheng2025diffusionnft} is detailed in the supplementary material. Given that existing reward models lack the capability to evaluate Chinese prompts, we employ a training dataset consisting exclusively of English prompts. During inference, all parameter settings align with official recommendations, and the random seed is fixed at 0.

As illustrated in the figures, PosterReward simultaneously optimizes aesthetic appeal and enhances both the layout quality and text-rendering accuracy of graphic designs. In contrast, existing reward models exhibit a noticeable performance gap compared to PosterReward, as they lack the specialized capacity to perceive and evaluate textual and structural layout elements.

\subsection{PosterBench}

\subsubsection{Benchmark Details}

We introduce \textbf{PosterBench}, a benchmark designed to evaluate the performance of various image generation models on poster creation tasks using PosterReward. PosterBench comprises 250 distinct poster generation prompts, of which 100 are related to cinematic themes and 150 cover other diverse categories. To assess generation stability, we instruct each model to generate eight posters per prompt. We evaluate six state-of-the-art closed-source models (Nano-Banana-Pro, Seedream-4.5*, Nano-Banana, Seedream-4.0, GPT-Image-1, and Seedream-3.0) and seven representative open-source models (Qwen-Image-2512, Qwen-Image, Z-Image-Turbo, Flux.2-klein-9B, Flux.1-krea-dev, Flux.1-dev, and SD3.5-L). Notably, for Seedream-4.5*, which enforces a minimum resolution constraint, we generate images at a target 2K resolution and subsequently resize them to the standard benchmark resolution for evaluation.

\begin{table}[!t]
\centering
    \setlength{\abovecaptionskip}{0.1cm} 
    \setlength{\belowcaptionskip}{-0cm}  
\captionsetup{font=small}
\caption{Performance of different models on PosterBench. Mean, Median, and Bo8-Avg represent the respective scores. Std-Avg denotes the mean of the standard deviations calculated for each group, indicating stability (lower is better).}
\label{tab:posterbench_results}
{\footnotesize
\renewcommand\arraystretch{1.2}
\begin{tabular*}{0.8\linewidth}{@{\extracolsep{\fill}}l cccc@{}}
\toprule
\textbf{Model} & \textbf{Mean} $\uparrow$ & \textbf{Median} $\uparrow$ & \textbf{Std-Avg} $\downarrow$ & \textbf{Bo8-Avg} $\uparrow$ \\
\midrule
\multicolumn{5}{@{}l}{\textit{Closed-Source Models}} \\
Nano-Banana-Pro  & \textbf{13.36} & \textbf{13.47} & \underline{1.91} & \textbf{15.77} \\
Seedream-4.5*     & \underline{12.03} & \underline{12.09} & 2.08 & \underline{14.57} \\
Nano-Banana\cite{google_gemini25_flash_image_2025} & 11.60 & 11.69 & 2.17 & 14.49 \\
Seedream-4.0\cite{seedream2025seedream4} & 11.46 & 11.44 & 2.06 & 13.93 \\
GPT-Image-1      & 11.16 & 11.38 & \textbf{1.75} & 13.43 \\
Seedream-3.0\cite{gao2025seedream3} & 5.01 & 5.13 & 3.66 & 9.75 \\
\midrule
\multicolumn{5}{@{}l}{\textit{Open-Source Models}} \\
Qwen-Image-2512\cite{wu2025qwenimagetechnicalreport} & \textbf{11.86} & \textbf{11.63} & \textbf{1.46} & \textbf{13.85} \\
Qwen-Image\cite{wu2025qwenimagetechnicalreport}      & \underline{7.69} & \underline{7.72} & 2.55 & 11.06 \\
Z-Image-Turbo\cite{cai2025zimage} & 7.65 & 7.31 & \underline{2.18} & 10.47 \\
Flux.2-klein-9B\cite{cai2025zimage} & 7.38 & 7.66 & 3.20 & \underline{11.67} \\
Flux.1-krea-dev\cite{flux1kreadev2025} & 5.00 & 5.14 & 3.59 & 9.58 \\
Flux.1-dev\cite{flux2024} & 2.55 & 2.42 & 3.85 & 7.81 \\
SD3.5-L\cite{esser2024scalingrectifiedflowtransformers} & -2.90 & -3.92 & 2.68 & 1.24 \\
\bottomrule
\end{tabular*}
}
\end{table}

\subsubsection{Evaluation Results}

We evaluate the generated samples from each model based on several key metrics. To measure the average generation quality, we compute the mean and median scores. Std-Avg denotes the mean of the standard deviations calculated for each group, indicating stability (lower is better). Finally, to evaluate the model's optimal performance across multiple trials, we use the Best-of-8 Average (Bo8-Avg) score, which is the mean of the highest-scoring sample from each set of eight generations.

As shown in Table~\ref{tab:posterbench_results}, Nano-Banana-Pro leads the overall benchmark, with Seedream-4.0 and 4.5 also demonstrating outstanding top-tier performance. While older closed-source models like Seedream 3.0 show limited capabilities in precise text rendering and layout generation, Qwen-Image-2512 establishes a dominant position in the open-source category. Notably, while a significant gap persists between most other open-source models (e.g., Qwen-Image, Z-Image-Turbo, and Flux.2-klein-9B) and leading closed-source solutions, Qwen-Image-2512 successfully achieve a performance level nearly on par with advanced closed-source models.

\section{Conclusion}

We introduce \textbf{PosterReward}, the first reward model capable of accurately assessing the quality of generated posters. We begin by constructing an automated data collection pipeline to curate Poster-Preference-70K, a high-quality dataset of poster preferences derived from a large volume of raw data. Subsequently, we employ a multi-stage cascaded framework that leverages image analysis to enhance the performance of our reward models. We also propose \textbf{PosterRewardBench} and \textbf{PosterBench}, which establish new benchmarks for measuring the performance of reward models on poster evaluation tasks and the poster generation quality of image synthesis models.

\clearpage

\bibliographystyle{plainnat}
\bibliography{main}

\clearpage
\setcounter{page}{1}

This is supplementary material for \textbf{\textit{PosterReward: Unlocking Accurate Evaluation for High-Quality Graphic Design Generation.}} 

We present the following materials in this supplementary material:

\begin{itemize}
    \item \textbf{Sec.\ref{test_time_scaling}} Qualitative comparison of Best-of-8 selection performance against HPSv3, UnifiedReward, and MLLMs, highlighting robustness to scoring failures and position bias.
    
    \item \textbf{Sec.\ref{rl_diffusion_nft}} Visual and quantitative comparison of SD3.5-Medium fine-tuned via Diffusion-NFT using PosterReward versus baselines like HPSv3, UnifiedReward, and PaddleOCR.
    
    \item \textbf{Sec.\ref{dataset_quality}} Analysis of the dataset construction pipeline, including multi-model verification, position bias mitigation strategies, and quality assessment via Kendall’s Coefficient of Concordance.
    
    \item \textbf{Sec.\ref{dataset_ablation}} Investigation into the impact of dataset scale, filtering stringency, and the integration of general aesthetic data on downstream model performance.
    
    \item \textbf{Sec.\ref{exp_setup}} Detailed training hyperparameters within the MS-Swift framework and construction specifications for the PosterRewardBench Advanced and Basic subsets.
    
    \item \textbf{Sec.\ref{limitation_future_work}} Discussion on computational overhead and MLLM evaluation constraints, alongside future directions for efficiency and scalability.
\end{itemize}

\section{Test-Time-Scaling with PosterReward via Best-of-8 Selection}
\label{test_time_scaling}

\begin{figure*}[ht]
    \centering
    \includegraphics[page=1, width=\textwidth]{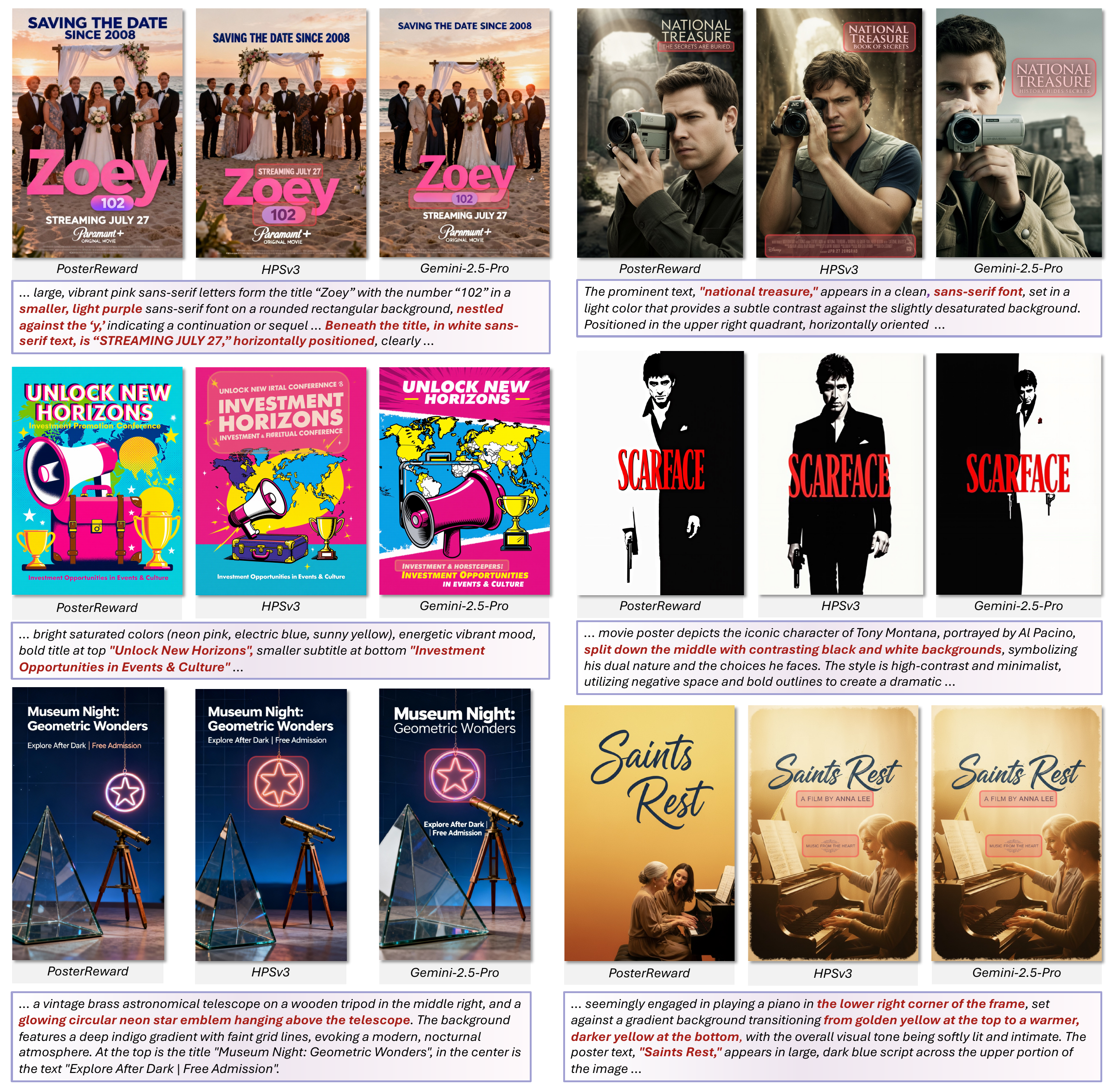}
    \caption{\textbf{Test-Time-Scaling results using Best-of-8 (Bo8) Selection.} Comparison of poster selection performance between PosterReward, HPSv3, and Gemini-2.5-Pro. PosterReward demonstrates superior accuracy in assessing text and layout quality, avoiding the scoring failures and position biases observed in other models.}
    \label{fig_bo8}
\end{figure*}

We investigate the feasibility of using PosterReward for poster selection. Based on the poster data generated by various models in the PosterBench benchmark, we selected several samples and apply Test-Time-Scaling via Best-of-8 (Bo8) selection. The results are presented in Figure~\ref{fig_bo8}.

As illustrated in the figure, PosterReward demonstrates a more accurate assessment of the poster content. Compared to HPSv3\cite{ma2025hpsv3}, PosterReward can more precisely evaluate text accuracy and layout requirements, showing a significant advantage in graphic design scenarios. UnifiedReward\cite{wang2025unifiedreward} is unsuitable for Bo8 selection, as it frequently produces identical scores for different candidates, making it difficult to distinguish the best one. We also attempted Bo8 selection using Multi-modal Large Language Models (MLLMs), which involved providing multiple images as input and prompting the model to identify the index of the optimal one. Our experiments indicate that the MLLMs struggle to make accurate choices among several visually similar images. Furthermore, we observe that even state-of-the-art proprietary MLLMs exhibit position bias\cite{shi2024judging,li2024split} when selecting among multiple similar images. Taking Gemini-2.5-Pro\cite{comanici2025gemini} as an example, when the first image in a sequence is of sufficient quality, the model demonstrates a tendency to select it, with its Chain-of-Thought (CoT) predominantly focusing on justifying the first image as the optimal candidate. The prompt for using MLLM for Bo8 selection is shown in Figure \ref{prompt:1-1-rank}.

\section{Reinforcement Learning with PosterReward via Diffusion-NFT}
\label{rl_diffusion_nft}

\begin{figure*}[ht]
    \centering
    \includegraphics[page=1, width=0.9\textwidth]{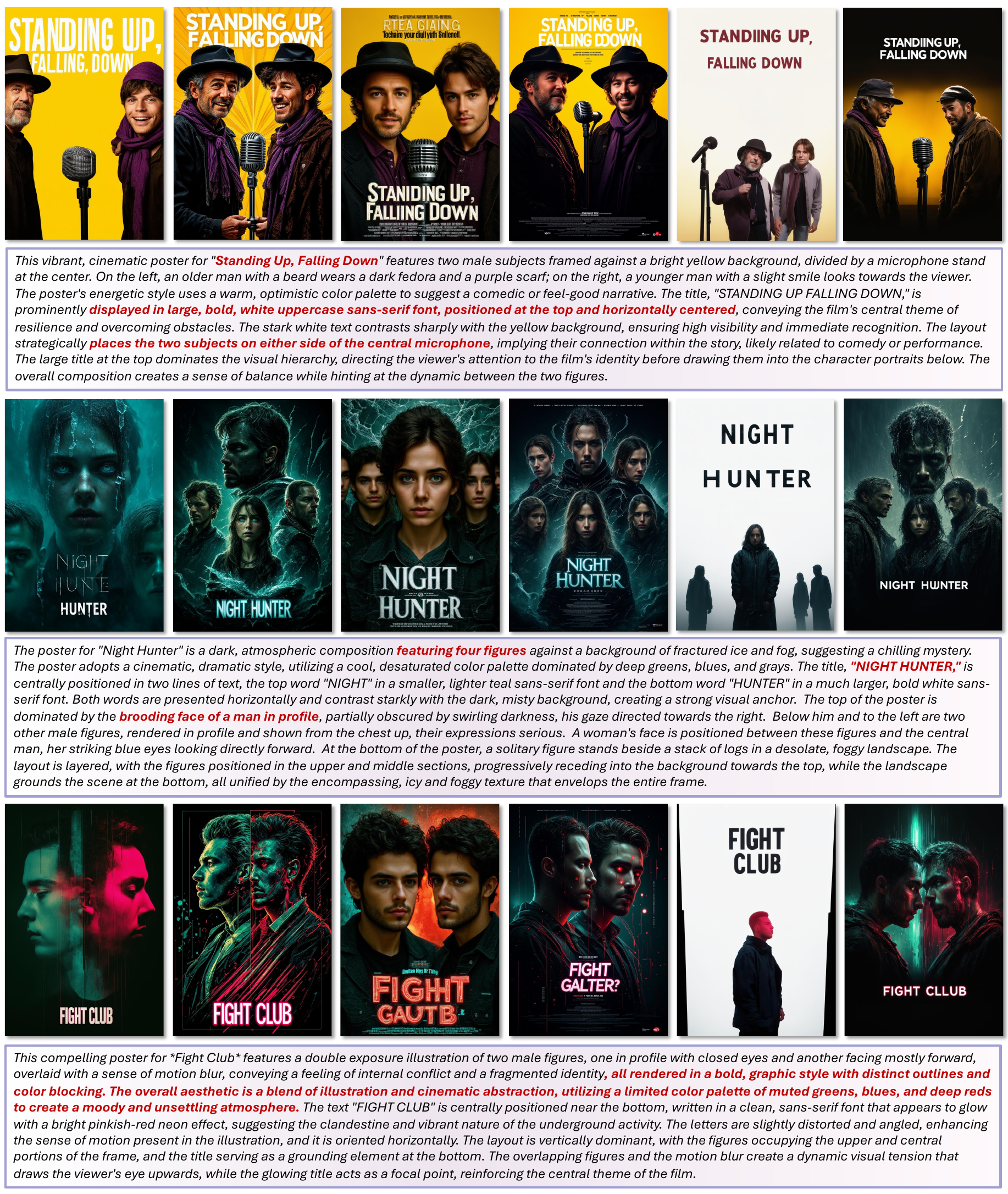}
    \caption{\textbf{Visual comparison of SD3.5-Medium fine-tuned with various reward models (Part 1).} From left to right, the columns display the outputs of the original SD3.5-Medium, followed by models fine-tuned using PosterReward, HPSv3, UnifiedReward, PaddleOCR, and the combined UnifiedReward + PaddleOCR. The corresponding prompts are enclosed in the purple boxes at the bottom.}
    \label{fig_nft_part1}
\end{figure*}

\begin{figure*}[ht]
    \centering
    \includegraphics[page=1, width=0.9\textwidth]{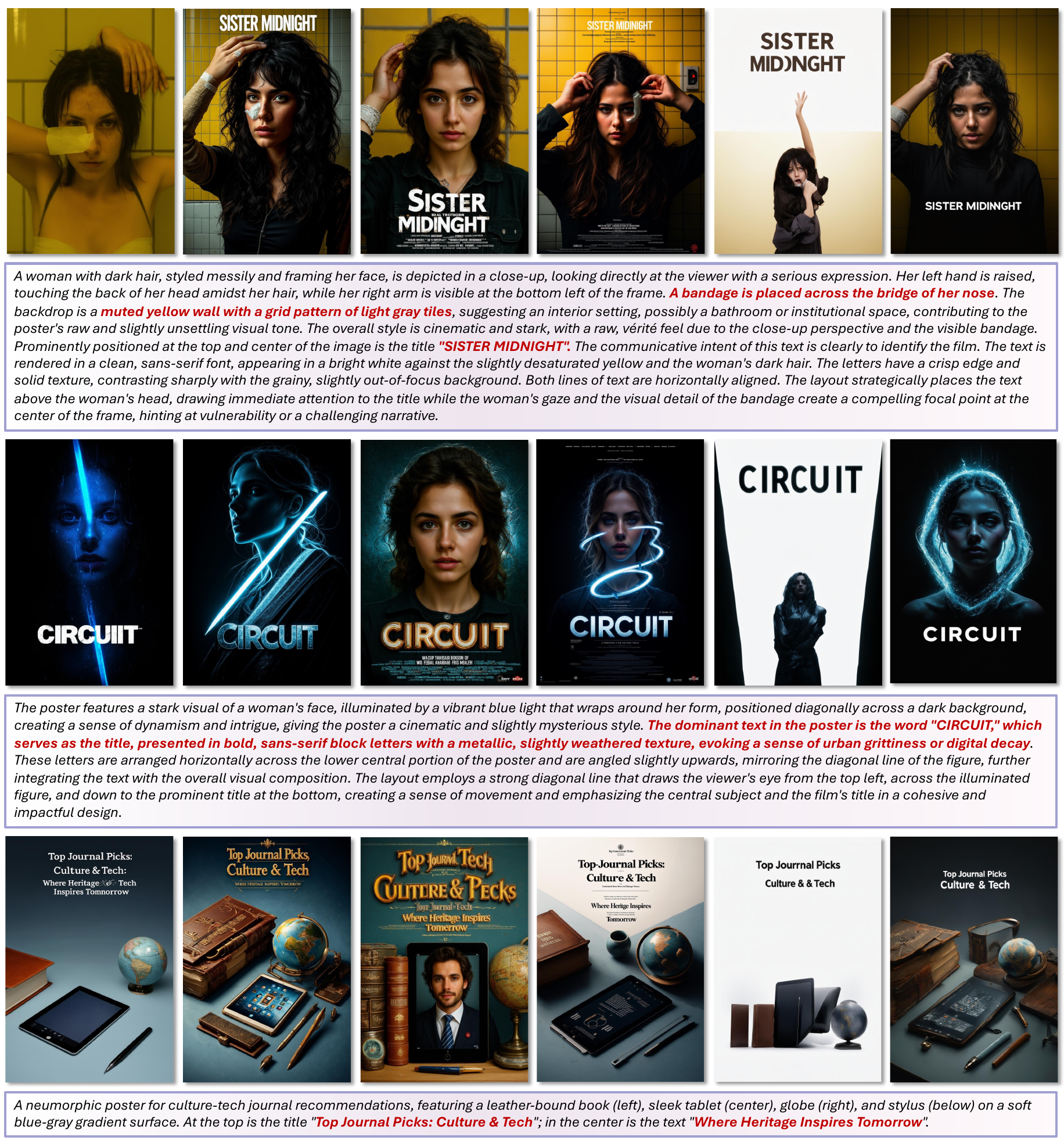}
    \caption{\textbf{Visual comparison of SD3.5-Medium fine-tuned with various reward models (Part 2).} Consistent with the previous figure, the columns represent the original SD3.5-Medium, PosterReward, HPSv3, UnifiedReward, PaddleOCR, and the joint UnifiedReward + PaddleOCR, respectively. Text prompts are provided in the bottom purple boxes.}
    \label{fig_nft_part2}
\end{figure*}

\begin{figure*}[ht]
    \centering
    \includegraphics[page=1, width=\textwidth]{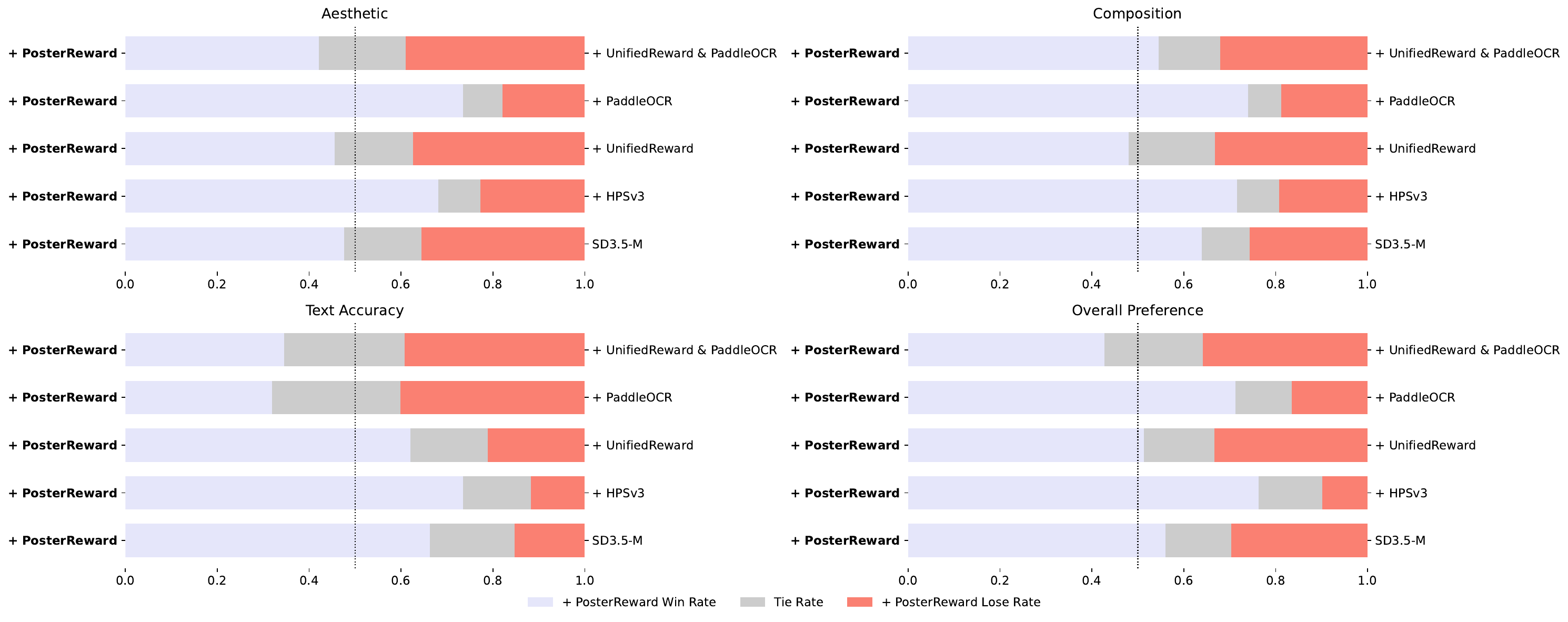}
    \caption{\textbf{User study results comparing PosterReward with varying baselines.} The evaluation covers four dimensions: Aesthetic, Composition, Text Accuracy, and Overall Preference. The bars represent the win (purple), tie (gray), and loss (red) rates of the model fine-tuned with PosterReward against SD3.5-M, HPSv3, UnifiedReward, and PaddleOCR\cite{cui2025paddleocr30technicalreport}. PosterReward demonstrates a consistent preference advantage across most metrics.}
    \label{fig_nft_user_study}
\end{figure*}

With the increasing prevalence of post-training reinforcement learning (RL) in image generation, numerous evaluation models have been adopted as reward models to furnish optimization signals. Consequently, we evaluate the efficacy of RL utilizing PosterReward. We select DiffusionNFT\cite{zheng2025diffusionnft} as our reinforcement learning paradigm. Unlike conventional Policy Gradient methods that discretize the reverse sampling process, DiffusionNFT optimizes the diffusion model directly on the forward process via the flow matching objective. It contrasts positive and negative generations to define an implicit policy improvement direction, naturally incorporating reinforcement signals into a supervised learning objective:
\begin{equation}
\begin{split}
    \mathcal{L}(\theta) = \mathbb{E}_{\boldsymbol{c}, \pi^{\text{old}}, t} \Big[ & r \|\boldsymbol{v}_\theta^+(\boldsymbol{x}_t, \boldsymbol{c}, t) - \boldsymbol{v}\|_2^2 \\
    & + (1-r) \|\boldsymbol{v}_\theta^-(\boldsymbol{x}_t, \boldsymbol{c}, t) - \boldsymbol{v}\|_2^2 \Big]
\end{split}
\end{equation}
where $\boldsymbol{v}_\theta^+$ and $\boldsymbol{v}_\theta^-$ denote the implicit positive and negative velocities, respectively. This formulation enables efficient optimization compatible with arbitrary black-box solvers without requiring likelihood estimation.

We benchmark against reward models including HPSv3\cite{ma2025hpsv3} and UnifiedReward-Qwen-7B. Both output scalar scores, representing state-of-the-art discriminative and generative reward paradigms, respectively. Furthermore, given our specific focus on poster and graphic design, we incorporate PaddleOCR as a reward model; it provides signals by evaluating the Levenshtein edit distance between the recognized text and the ground-truth text required for rendering. We employ Stable Diffusion 3.5 Medium (SD3.5-M) as the base model. The fine-tuning process utilizes 8 NVIDIA A100 GPUs, with an additional 4 A100 GPUs dedicated to reward model deployment. All training hyperparameters adhere to the original DiffusionNFT configuration. We select the model fine-tuned for 350 steps for evaluation. For the original SD3.5-M model, we followed the setting of CFG=4.5, while for the fine-tuned model, we used CFG=1.0. The seed used during generation is always fixed at 0.

We present a comprehensive evaluation of SD3.5-M optimized by various reward models, with visual comparisons detailed in Figure~\ref{fig_nft_part1} and Figure~\ref{fig_nft_part2}, and quantitative user study results summarized in Figure~\ref{fig_nft_user_study}. Regarding \textbf{text accuracy}, although HPSv3\cite{ma2025hpsv3} and UnifiedReward\cite{wang2025unifiedreward} effectively improve the rendering of key headlines, they inadvertently introduce a substantial amount of illegible small text—resembling ``credit blocks''—which severely compromises the overall visual coherence. In contrast, PosterReward significantly elevates text rendering quality while effectively preventing the generation of redundant content. Furthermore, while incorporating PaddleOCR\cite{cui2025paddleocr30technicalreport} as a fine-tuning signal yields higher character recognition accuracy, it compromises textual aesthetics (e.g., font style and integration) and exhibits a lower success rate when handling complex, multi-text rendering scenarios. In terms of \textbf{aesthetics and composition}, PosterReward provides superior and more precise reward signals, whereas HPSv3 underperforms in both dimensions. While UnifiedReward exhibits a relatively high standard of aesthetics and composition, its scores are diminished by text accuracy defects and a noticeable reduction in image realism. Additionally, the dual-reward model combining PaddleOCR and UnifiedReward marginally sacrifices aesthetic and compositional quality to prioritize text correctness. Consequently, in terms of \textbf{overall preference}, the model fine-tuned with PosterReward successfully outperforms all competitors, demonstrating its capability to provide high-quality, holistic reward signals for poster generation tasks.

\section{Details and Quality Analysis of the AI Preference Dataset}
\label{dataset_quality}

In the preference data pipeline, we employ four Multimodal Large Language Models (MLLMs): GLM-4.5v\cite{vteam2025glm45vglm41vthinkingversatilemultimodal}, GPT-5, Gemini-2.5-Flash-lite, and Gemini-2.5-Pro. The ``Thinking'' mode is activated for all models; specifically, for the Gemini-2.5-Flash-lite and Gemini-2.5-Pro, the thinking budget is set to 4096. Regarding sampling parameters, the temperature is set to 1.0 for GPT-5, while a value of 0.7 is maintained for the remaining models. The specific prompts utilized for ranking and preference judgment are presented in Figure~\ref{prompt:ranking-collection-p1}, Figure~\ref{prompt:ranking-collection-p2} and Figure~\ref{prompt:3-1-pairwise-preference}.

We employ Gemini-2.5-Pro to annotate data for both single-image analysis and the collection of paired-image preference Chain-of-Thought (CoT) sequences required for fine-tuning. The detailed prompts utilized for these tasks are presented in Figure~\ref{prompt:pointwise-collection} and Figure~\ref{prompt:pairwise-collection}. Specifically, during the acquisition of preference CoTs, we explicitly provide the ground-truth preference labels to the model, instructing it to articulate the reasoning behind the selection.

To evaluate the visual quality and semantic alignment of the generated movie posters, we employed a multi-round ranking strategy utilizing the Gemini-2.5-Flash-lite multimodal large language model. Given the stochastic nature of Large Language Models (LLMs), a single inference may not accurately reflect the model's stable preference. Therefore, for each creative brief, we presented a set of $m=6$ candidate images to the model and instructed it to rank them from best to worst based on specific criteria (e.g., prompt adherence, aesthetic quality, and text rendering). This process was repeated $k=6$ times independently to obtain a distribution of rankings.

To quantify the reliability and consistency of the model's rankings across these $k$ iterations, we calculated Kendall's Coefficient of Concordance (Kendall's $W$)\cite{kendall1939problem}. Kendall's $W$ is a non-parametric statistic that assesses agreement among raters—in this context, the distinct inference rounds acts as pseudo-raters. 

Let $m$ be the number of images (items) and $k$ be the number of ranking rounds (raters). Let $r_{i,j}$ denote the rank assigned to the $i$-th image in the $j$-th round, where $1 \le r_{i,j} \le m$. The sum of ranks, $R_i$, given to the $i$-th image is calculated as:

\begin{equation}
    R_i = \sum_{j=1}^{k} r_{i,j}
\end{equation}

The mean of the sum of ranks, $\bar{R}$, is given by:

\begin{equation}
    \bar{R} = \frac{1}{m} \sum_{i=1}^{m} R_i = \frac{k(m+1)}{2}
\end{equation}

Kendall's $W$ is defined as the ratio of the observed sum of squared deviations ($S$) to the maximum possible sum of squared deviations. First, we calculate $S$:

\begin{equation}
    S = \sum_{i=1}^{m} (R_i - \bar{R})^2
\end{equation}

Assuming no ties in the rankings, Kendall's $W$ is computed as:

\begin{equation}
    W = \frac{12S}{k^2(m^3 - m)}
\end{equation}

The value of $W$ ranges from 0 to 1, where $W=1$ indicates complete agreement across all ranking rounds (implying the model's preference is highly stable), and $W=0$ indicates no agreement. We utilize this metric to filter out instances where the model hallucinates or fails to distinguish quality differences effectively, thereby ensuring the robustness of the final averaged ranking.

To further guarantee the robustness of the constructed preference pairs and mitigate the potential biases inherent to a single evaluator, we implemented a rigorous multi-model verification pipeline. We employed an ensemble of three state-of-the-art multimodal large language models: Gemini-2.5-Pro, GPT-5, and GLM-4.5v. 

Addressing the phenomenon of position bias—where models may exhibit a systematic preference for the first or second image in a sequence—we adopted a bidirectional evaluation strategy. For every candidate pair $(I_A, I_B)$, each of the three models performed the evaluation twice: once with the original order and once with the image order swapped (i.e., $(I_B, I_A)$). This procedure yields a total of six independent assessments per sample. In each assessment, the models were tasked with a holistic evaluation covering fundamental integrity, typographical precision, and artistic quality, classifying the relationship as a clear preference, a tie, or a rejection of both images. This ensemble approach allows us to quantify the confidence level of each preference label based on the consensus among the six judgments.

\begin{table}[t]
\centering
\captionsetup{font=small}
\caption{Analysis of dataset quality under different filtering criteria. N denotes the number of samples. Corr., Err., and Controv. represent the percentages of samples classified as Correct, Error, and Controversial, respectively.}
\label{tab:filtering_analysis_condensed}
\setlength{\tabcolsep}{5pt} 
{\footnotesize 
\renewcommand\arraystretch{1.2} 
\begin{tabular}{@{}l cccc@{}}
\toprule
\textbf{Method} & \textbf{N} & \textbf{Corr. (\%)} & \textbf{Err. (\%)} & \textbf{Controv. (\%)} \\
\midrule
All Correct             & 20K & 87.2 & 4.6 & 8.2  \\
5 Correct + 1 Tie       & 33K & 85.0 & 5.8 & 9.2  \\
5 Correct + 1 Tie/Error & 70K & 78.6 & 10.7 & 10.7  \\
\bottomrule
\end{tabular}
}
\end{table}

To compare the reliability of our automated annotation framework under different filtering conditions, we manually annotate and analyze a subset of 1,000 samples from PosterPreference-70K. As described in Section 3, the final filtering stage involves three models, each performing two evaluations with the image order swapped, resulting in a total of six assessments per sample. We investigate three filtering strategies with varying levels of stringency: (1) retaining samples where all six model assessments are unanimous; (2) retaining samples with at least five consistent preference assessments, allowing for one tie; and (3) retaining samples with unanimous preference, allowing for one tie or one conflicting assessment. These samples are then annotated by four human experts and categorized into three groups: \textit{Correct}, where the preference aligns with at least three annotators; \textit{Error}, where the preference contradicts at least three annotators; and \textit{Controversial} for all other cases.


As shown in Tab.~\ref{tab:filtering_analysis_condensed}, while the consistency between the automated filtering and human evaluations naturally decreases as the criteria become less stringent, the \textit{5 Correct + 1 Tie/Error} strategy still maintains a robust alignment rate of 78.6\%, with a relatively low error rate of 10.7\%. This indicates that even under a slightly more relaxed filtering threshold, the data quality remains within an acceptable range for effective preference learning. Consequently, we elect to utilize the full 70K dataset for training. We posit that the significant increase in sample diversity and volume (from 20K/33K to 70K) outweighs the marginal gain in precision offered by stricter filtering, as larger-scale datasets typically drive better generalization capabilities in reward modeling. This 70K dataset is subsequently combined with a 100K preference pair subset from HPDv3 to form our final training corpus.

\section{Ablation Studies on the Dataset Components}
\label{dataset_ablation}

\begin{table}[ht]
\centering
    \setlength{\abovecaptionskip}{0.1cm} 
    \captionsetup{font=small}
\caption{Ablation study on dataset components. We compare the impact of using partial (33K) vs. full (70K) PosterPreference (PP) data, both independently and in combination with HPDv3. All values represent accuracy ($\uparrow$).}
\label{tab:dataset_ablation}
\setlength{\tabcolsep}{5pt} 
{\footnotesize
\renewcommand\arraystretch{1.2}
\begin{tabular}{@{}l ccc@{}}
\toprule
\textbf{Training Dataset Combination} & \textbf{HPDv3} & \textbf{PRB-Ad} & \textbf{PRB-Basic} \\
\midrule
PosterPreference-33K (Ours)        & 63.0 & 84.9 & 81.9 \\
PosterPreference-70K (Ours)        & 63.9 & 84.6 & 83.0 \\
HPDv3-100K                         & 75.8 & 34.1 & 57.0 \\
HPDv3 + PP-33K                     & 76.9 & 84.5 & 80.8 \\
\textbf{HPDv3 + PP-70K (Final)}    & \textbf{77.1} & \textbf{85.0} & \textbf{83.9} \\
\bottomrule
\end{tabular}
}
\end{table}

To validate our dataset selection strategy and quantify the contribution of different data sources, we conducted comprehensive ablation studies. Specifically, we investigate the performance of reward models trained on five distinct dataset configurations: 
(1) The high-consistency subset of our domain-specific dataset (\textbf{PP-33K}); 
(2) The full filtered domain-specific dataset (\textbf{PP-70K}); 
(3) The general aesthetic preference dataset (\textbf{HPDv3-100K}); 
(4) A combination of HPDv3 and the high-consistency subset (\textbf{HPDv3 + PP-33K}); 
and (5) A combination of HPDv3 and the full domain dataset (\textbf{HPDv3 + PP-70K}). We evaluated these models on the general HPDv3 test set as well as our domain-specific benchmarks, PRB-Ad and PRB-Basic.

As presented in Table~\ref{tab:dataset_ablation}, models trained solely on domain-specific data (PP-33K/70K) perform well on poster-related benchmarks but show limited generalization on the general HPDv3 test set. Conversely, training exclusively on HPDv3 yields strong general aesthetic judgment but suboptimal performance on poster-specific tasks (PRB-Ad/Basic), indicating a domain gap.

Crucially, combining the general and domain-specific datasets yields significant performance gains. Comparing the two combined settings, the model trained with the full \textbf{PP-70K} dataset consistently outperforms the one trained with the smaller \textbf{PP-33K} subset across all benchmarks. We also noted that the performance improvement was particularly significant on the out-of-domain test set PRB-Basic, indicating that more samples are beneficial for improving the generalization ability of the reward model. This empirical evidence supports our hypothesis in Section 3 that the benefits of increased data scale and diversity in the 70K dataset outweigh the marginal improvements in label consistency found in the 33K subset. Consequently, the \textit{HPDv3 + PP-70K} configuration is adopted as our final training set, achieving the best balance between general aesthetic understanding and domain-specific expertise.

\begin{figure*}[ht]
    \centering
    \includegraphics[page=1, width=\textwidth]{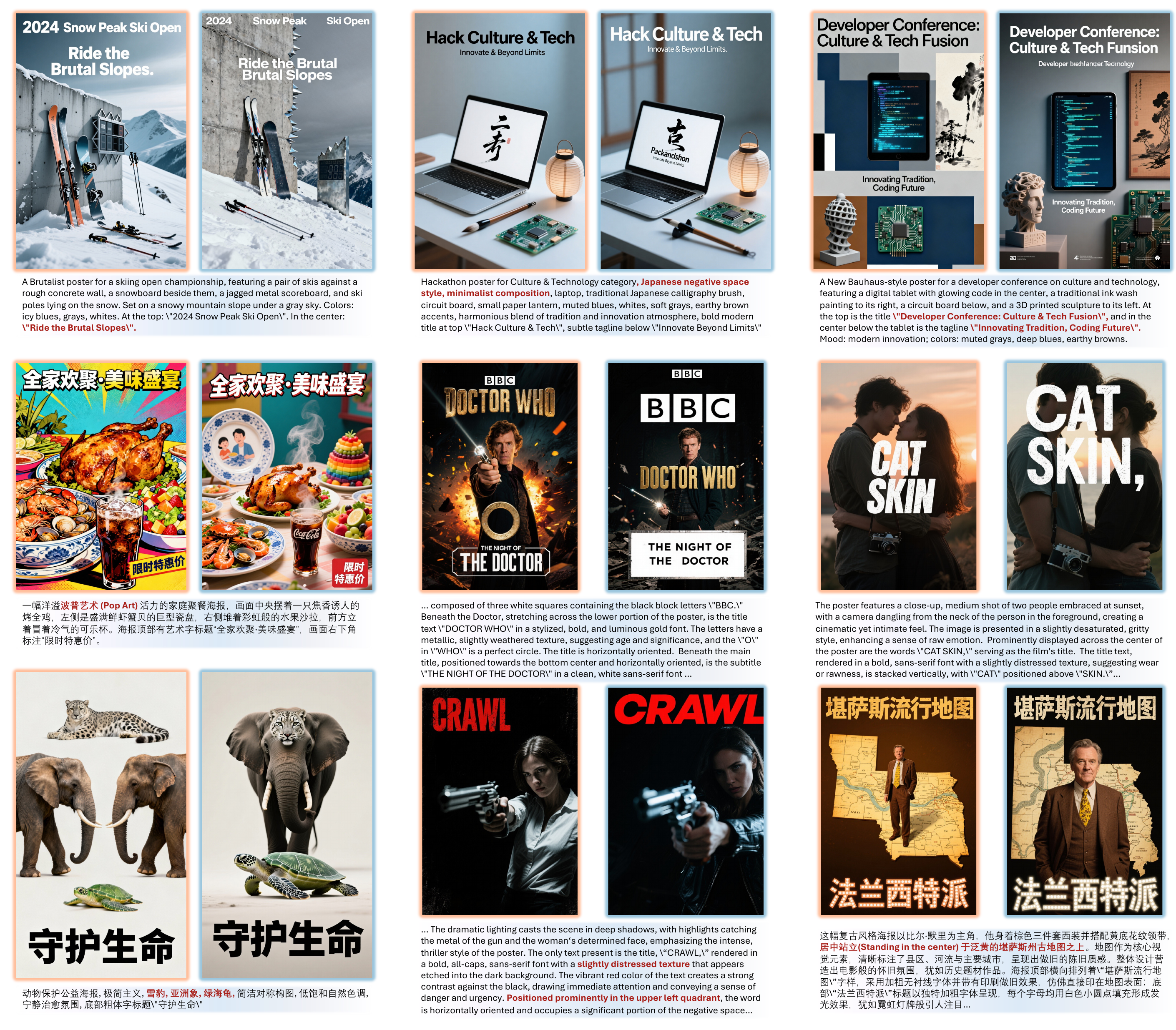}
    \caption{\textbf{Sample preference pairs from PosterRewardBench-Advanced.} This subset comprises images generated by Seedream and Qwen-Image-Lighting. It features complex poster layouts and supports both Chinese and English text rendering evaluations. The left side of each group shows the chosen image, the right side shows the rejected image, and the bottom displays an excerpt from the prompt.}
    \label{fig_prb_ad}
\end{figure*}

\begin{figure*}[ht]
    \centering
    \includegraphics[page=1, width=\textwidth]{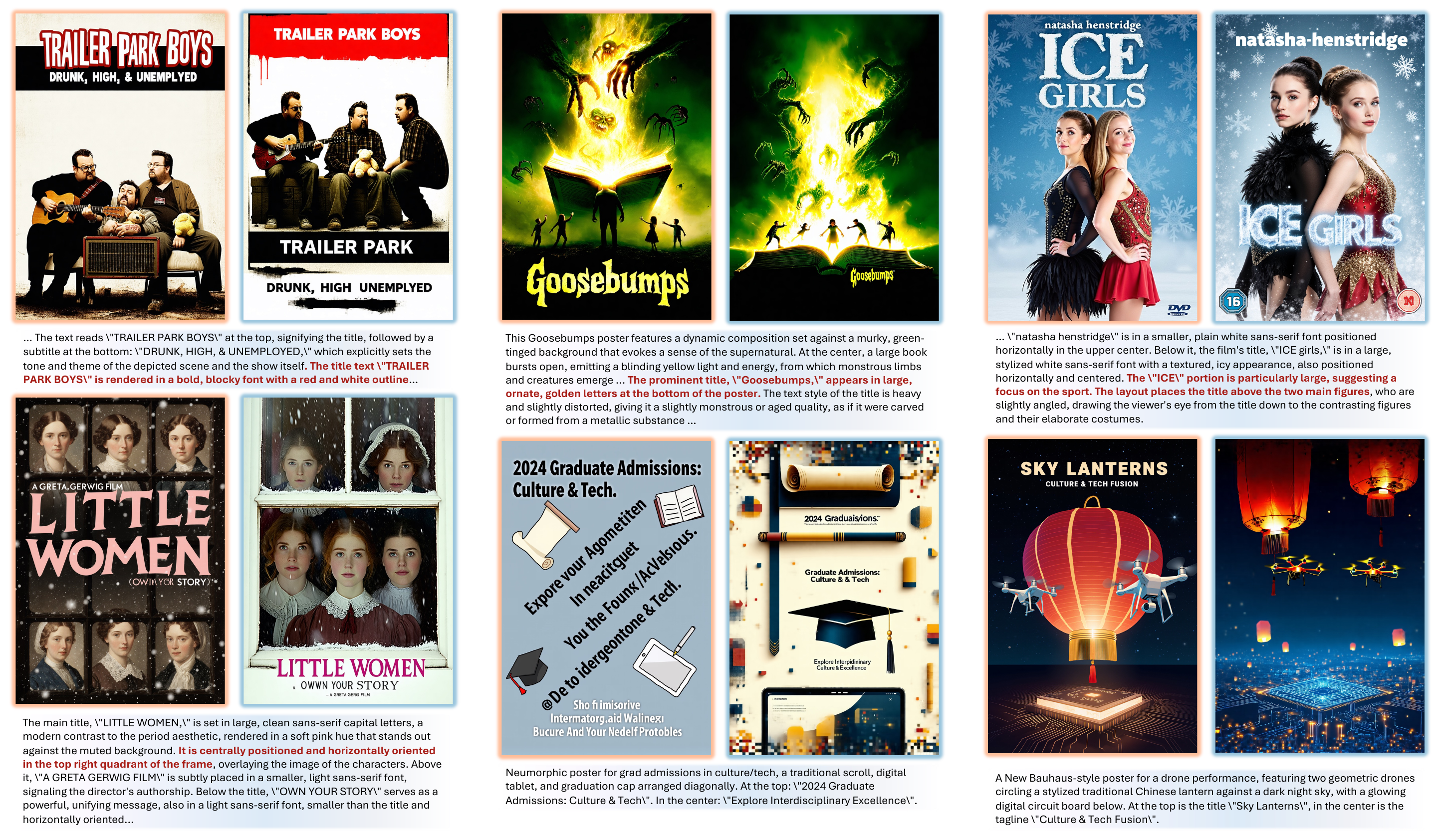}
    \caption{\textbf{Sample preference pairs from PosterRewardBench-Basic.} This subset consists of images generated by Flux.1 and SD3.5 models. Due to model limitations, this benchmark focuses exclusively on English text rendering and standard graphic designs. The left side of each group shows the chosen image, the right side shows the rejected image, and the bottom displays an excerpt from the prompt.}
    \label{fig_prb_basic}
\end{figure*}

\section{Detailed Experimental Settings, Benchmark Construction}
\label{exp_setup}

All experimental stages are conducted within the MS-Swift framework. We utilize DeepSpeed Zero-3 for Joint Supervised Fine-Tuning (Stage 1), Joint Rejection Sampling Fine-Tuning (Stage 2), and GRPO Fine-Tuning (Stage 4), while Zero-2 is employed for Scoring Module Training (Stage 3). Regarding hyperparameters, we set the per-device batch size to 1 with 2 gradient accumulation steps for the first two stages. For Stage 3, the batch size is set to 4 with 2 accumulation steps, whereas the GRPO stage uses a batch size of 1 with 8 accumulation steps. The adamw\_torch optimizer is consistently applied across all stages.

\noindent{\textbf{PosterRewardBench.}} The PosterRewardBench comprises a total of 1,740 preference pairs. Specifically, PosterRewardBench-Advanced contains 1,223 pairs derived from Seedream 3.0, Seedream 4.0, and Qwen-Image-Lighting, while PosterRewardBench-Basic consists of 517 pairs sourced from Flux.1-dev, Flux.1-Krea-dev, and SD3.5-M. Within the Advanced subset, 488 pairs ($39.9\%$) feature Chinese prompts and font rendering instructions, with the remainder comprising English counterparts. Conversely, given the incompatibility of Flux and SD3.5 with the Chinese language, PosterRewardBench-Basic consists exclusively of English prompts. These pairs have been filtered from initial pools of 1,500 and 600 raw pairs, respectively. Four professional annotators evaluate preferences across multiple dimensions, and we retain only those pairs where at least three annotators reach a consensus. The data collection process spans approximately eight working days, with sample pairs illustrated in Figure~\ref{fig_prb_ad} and Figure~\ref{fig_prb_basic}. In the pairwise evaluation, to verify decision stability regarding input order, we perform assessments with the ``Chosen'' image positioned both first and last. The specific prompts utilized are detailed in Figure~\ref{prompt:pairwise-prediction}. We also provide the prompt used by the analysis module when performing single-image analysis, as shown in Figure~\ref{prompt:pointwise-generation}.

\noindent{\textbf{PosterBench.}} The PosterBench evaluation dataset consists of 250 prompts randomly sampled from PosterRewardBench, stratified into 100 cinematic and 150 non-cinematic themes. To rigorously assess compositional generalization, we employ distinct resolution strategies for each category. All cinematic prompts are standardized to a fixed vertical resolution of $832 \times 1216$ pixels (approximate aspect ratio of $2:3$), reflecting conventional movie poster standards. Conversely, the non-cinematic subset is designed to test adaptability to diverse geometries; each of the 150 prompts is assigned a unique target resolution $(H_i, W_i)$ derived from its corresponding reference ground truth. Crucially, while target aspect ratios vary significantly \textit{between} different prompts—spanning landscape, portrait, and square formats—the target resolution for any \textit{specific} prompt remains constant across all evaluated models, ensuring a consistent baseline for comparative analysis.

Implementation of these target resolutions varies by model accessibility. Open-weights models (e.g., Qwen-Image\cite{qwenimage}, Flux\cite{flux2024}) are conditioned to generate images at the exact target dimensions $(H_i, W_i)$ without resizing. However, for proprietary APIs with discrete resolution constraints, we employ an adaptive mapping strategy to approximate the target geometry. For Nano-Banana\cite{google_gemini25_flash_image_2025}, arbitrary target dimensions are mapped to the nearest supported aspect ratio (e.g., $1:1, 2:3, 16:9$). Similarly, for GPT-Image-1, targets are bucketed into three presets: portrait ($1024 \times 1536$) if the aspect ratio $\alpha = W/H < 0.8$, landscape ($1536 \times 1024$) if $\alpha > 1.2$, and square ($1024 \times 1024$) otherwise. This approach minimizes distortion while ensuring that closed-source models adhere

\section{Limitation and Future Work}
\label{limitation_future_work}

Despite the promising results achieved by our framework, several limitations remain to be addressed.

First, regarding the inherent position bias in MLLMs, we currently mitigate this issue in pairwise comparisons through a data balancing strategy. However, this strategy is not easily scalable to multi-image evaluation scenarios (e.g., ranking a batch of images simultaneously), where the combinatorial complexity of position swapping becomes prohibitive. Future work will focus on exploring more robust mechanisms to enhance the reliability and consistency of MLLMs in multi-image assessment tasks, reducing the dependency on permutation-based data augmentation.

Second, our proposed two-stage reward model, while offering superior interpretability and accuracy, entails a higher computational cost compared to standard discriminative reward models. The requirement for two inference steps inherently limits the model's real-time inference efficiency. In future research, we aim to optimize this architectural pipeline to reduce computational overhead. Additionally, we plan to conduct a comprehensive analysis of model scaling to investigate how different model sizes impact the trade-off between performance and efficiency in multimodal preference learning.

\begin{figure*}[!t]
    \setlength{\abovecaptionskip}{0.1cm} 
    \setlength{\belowcaptionskip}{0cm}
\noindent\begin{example}{MLLM Bo8 Prompt}
The following eight images were all generated by this Prompt:

\textit{\{prompt\}}

Please compare these images and choose the one you think is the best to submit. Only the image number (1-8) will be returned.
\end{example}
\caption{\textbf{Prompt} for MLLM Ranking in Best-of-8 Selection.}
\label{prompt:1-1-rank}
\end{figure*}

\begin{figure*}[ht]
    \centering
    \begin{example}{Pointwise Analysis Collection Prompt}
        \small
        You are a meticulous AI Image Quality Analyst and Creative Director. Your mission is to provide a detailed, structured analysis for a provided image based on the prompt.

        \textbf{Your Step-by-Step Analysis Process:}

        \textbf{1. Deconstruct the prompt:} \\
        First, carefully read and internalize every detail of the provided prompt. Isolate its core components to establish your evaluation criteria.

        \textbf{2. Structured Image-by-Image Analysis:} \\
        For \textbf{the provided image}, you will conduct a methodical evaluation based on the following five criteria. Your analysis \textbf{MUST} address all five points in this specific order, analyzing both compliance and deviation within each point.

        \begin{itemize}[leftmargin=*, nosep]
            \item \textbf{1. Fundamental Image Integrity:} Assess the image's foundational technical quality, independent of the prompt's creative instructions. Scrutinize for any objective flaws such as unintended blur, pixelation, overexposure, underexposure, or digital noise that detract from a professional finish.
            \item \textbf{2. AI Artifact and Realism Evaluation:} Examine the image specifically for common AI generation artifacts and assess its overall textural fidelity. If characters are present, conduct a rigorous check for anatomical inconsistencies, such as malformed hands, distorted facial features, or unnatural limb positioning. Evaluate whether the image achieves a believable, cohesive sense of realism or if artifacts disrupt the illusion.
            \item \textbf{3. Typographical Precision Analysis:} When the prompt includes text, perform a granular analysis of its execution. Evaluate adherence to the specified content, checking for spelling errors, omissions, or additions. Scrutinize the rendering of typographical details: font family, style, color, weight, and capitalization. Assess layout aspects like kerning, leading, alignment, and scale. Conclude by evaluating the overall legibility and seamless integration of the text.
            \item \textbf{4. Visual Prompt Interpretation:} Beyond text, assess the image's faithfulness to all other thematic and compositional directives in the prompt. Evaluate the accuracy of core elements (characters, objects, setting), the composition, layout, and adherence to the specified artistic style, mood, and color palette. Identify every element that aligns with the brief, as well as any deviations or creative misinterpretations.
            \item \textbf{5. Standalone Artistic Evaluation:} Disregarding the prompt's specific constraints, judge the image purely on its own artistic and aesthetic merits. Evaluate its composition, use of light and shadow, color theory, emotional impact, and overall visual appeal. Assess the technical execution and the creative choices made, determining if it is a compelling and well-crafted image in its own right.
        \end{itemize}

        \textbf{The prompt is as follows:} \\
        ``\{creative\_brief\}''

        \textbf{Required Output:} \\
        Your final output must be structured as follows. Do not include any conversational introductions or summaries outside of this defined structure.
        Provide your structured analysis for the image. Use a clear heading for \textbf{Image Analysis}. Within this analysis, you must use the five numbered subheadings in the specified order.
        The total length should be within 500 words.

        \textbf{(Example of the final part of the output)} \\
        \textbf{Image Analysis} \\
        \textbf{1. Fundamental Image Integrity:} \\
        \textnormal{[Your analysis here.]} \\
        ... \\
        \textbf{5. Standalone Artistic Evaluation:} \\
        \textnormal{[Your analysis here.]} \\
    \end{example}
    \caption{\textbf{Prompt used for Pointwise Analysis Data Collection.}}
    \label{prompt:pointwise-collection}
\end{figure*}

\begin{figure*}[h]
    \centering
    \begin{example}{Pairwise Preference Reasoning Prompt}
        \small
        The images were generated by the prompt: ``\{prompt\}''.

        \vspace{0.5em}
        Why is Image \{better\_image\_index\} better than Image \{worse\_image\_index\}?

        \vspace{0.5em}
        When analyzing, you can consider these dimensions: Image Quality, AI Artifacts, Prompt Adherence, Text Rendering, and Aesthetic Value. Focus only on the key differentiating factors that make Image \{better\_image\_index\} superior. You don't need to cover all dimensions—only explain the core reasons.

        \vspace{0.5em}
        Provide your analysis as a single paragraph.
    \end{example}
    \caption{\textbf{Prompt used for Pairwise Preference Reasoning Data Collection.}}
    \label{prompt:pairwise-collection}
\end{figure*}

\begin{figure*}[ht]
    \setlength{\abovecaptionskip}{0.1cm}
    \setlength{\belowcaptionskip}{0cm}
    
    \begin{example}{Pointwise Analysis Generation Prompt}
        \small
        Please analyze this image generated from the prompt: ``\{creative\_brief\}''. \\
        Provide a detailed analysis across these five dimensions: \\
        \textbf{1. Fundamental Image Integrity,} \\
        \textbf{2. AI Artifact and Realism Evaluation,} \\
        \textbf{3. Typographical Precision Analysis,} \\
        \textbf{4. Visual Prompt Interpretation, and} \\
        \textbf{5. Standalone Artistic Evaluation.}
    \end{example}
    \caption{\textbf{Prompt used for Pointwise Analysis Generation Training and Inference.}}
    \label{prompt:pointwise-generation}
\end{figure*}

\begin{figure*}[ht]
    \setlength{\abovecaptionskip}{0.1cm}
    \setlength{\belowcaptionskip}{0cm}
    
    \begin{example}{Pairwise Preference Prediction Prompt}
        \small
        The following two images are generated from this prompt: ``\{prompt\}''. \\
        \textbf{Is Image 1 better than Image 2?} \\
        Please answer \textbf{Yes} or \textbf{No} first, then provide the reason.
    \end{example}
    \caption{\textbf{Prompt used for Pairwise Preference Prediction Training and Inference.}}
    \label{prompt:pairwise-prediction}
\end{figure*}

\begin{figure*}[ht]
    \centering
    \begin{example}{Multi-Round Ranking Prompt (Part 1)}
        \small
        You are a meticulous AI Image Quality Analyst and Creative Director. Your mission is to provide a detailed, structured analysis for \{image\_ref\} based on a creative brief, culminating in a definitive ranking presented in JSON format.

        \vspace{0.5em}
        \textbf{Your Step-by-Step Analysis Process:}

        \vspace{0.5em}
        \textbf{1. Deconstruct the Creative Brief:} \\
        First, carefully read and internalize every detail of the provided prompt. Isolate its core components to establish your evaluation criteria.

        \vspace{0.5em}
        \textbf{2. Structured Image-by-Image Analysis:} \\
        For each of the \{num\_images\} images, you will conduct a methodical evaluation based on the following five criteria. Your analysis for each image \textbf{MUST} address all five points in this specific order, analyzing both compliance and deviation within each point.

        \vspace{0.5em}
        \textbf{Crucially, your analysis for each image must be completely independent and self-contained. Do not make comparisons or references to any other image within an individual analysis block (e.g., in the analysis for Image 1, do not mention Image 2). All comparative logic is reserved for the final ranking synthesis.}

        \vspace{0.5em}
        \begin{itemize}[leftmargin=*, nosep]
            \item \textbf{1. Fundamental Image Integrity:} Assess the image's foundational technical quality, independent of the prompt's creative instructions. Scrutinize for any objective flaws such as unintended blur, pixelation, overexposure, underexposure, or digital noise that detract from a professional finish. Conversely, note the image's strengths in clarity, sharpness, and clean rendering.
            \vspace{0.3em}
            \item \textbf{2. AI Artifact and Realism Evaluation:} Examine the image specifically for common AI generation artifacts and assess its overall textural fidelity. If characters are present, conduct a rigorous check for anatomical inconsistencies, such as malformed hands, distorted facial features, or unnatural limb positioning. Evaluate whether the image achieves a believable, cohesive sense of realism or if artifacts disrupt the illusion.
            \vspace{0.3em}
            \item \textbf{3. Typographical Precision Analysis:} When the prompt includes text, perform a granular analysis of its execution. Evaluate adherence to the specified content, checking for spelling errors, omissions, or additions. Scrutinize the rendering of typographical details: font family, style, color, weight, and capitalization. Assess layout aspects like kerning, leading, alignment, and scale. Conclude by evaluating the overall legibility and seamless integration of the text.
            \vspace{0.3em}
            \item \textbf{4. Visual Prompt Interpretation:} Beyond text, assess the image's faithfulness to all other thematic and compositional directives in the prompt. Evaluate the accuracy of core elements (characters, objects, setting), the composition, layout, and adherence to the specified artistic style, mood, and color palette. Identify every element that aligns with the brief, as well as any deviations or creative misinterpretations.
            \vspace{0.3em}
            \item \textbf{5. Standalone Artistic Evaluation:} Disregarding the prompt's specific constraints, judge the image purely on its own artistic and aesthetic merits. Evaluate its composition, use of light and shadow, color theory, emotional impact, and overall visual appeal. Assess the technical execution and the creative choices made, determining if it is a compelling and well-crafted image in its own right.
        \end{itemize}
    \end{example}
    \caption{\textbf{Prompt used for Multi-Round Ranking Data Collection (Part 1: Analysis Criteria).}}
    \label{prompt:ranking-collection-p1}
\end{figure*}

\begin{figure*}[p]
    \centering
    \begin{example}{Multi-Round Ranking Prompt (Part 2)}
        \small
        \textbf{3. Synthesize and Establish Ranking Logic:} \\
        After analyzing all images, synthesize your findings from the five-point analysis to establish a final ranking from best to worst. Your ranking \textbf{must} be a direct result of weighing your findings against this strict, \textbf{four-tier hierarchy of importance}:

        \vspace{0.5em}
        \begin{itemize}[leftmargin=*, nosep]
            \item \textbf{Priority 1: Fundamental Image Integrity.} This is the primary gatekeeper for quality. An image must first be technically sound. Any image with significant fundamental flaws (e.g., pervasive blur, noise, or exposure issues) will be penalized heavily, regardless of its performance in other areas. This is evaluated in \textbf{point 1}.
            \vspace{0.3em}
            \item \textbf{Priority 2: Comprehensive Prompt Adherence.} Once an image passes the fundamental quality check, its faithfulness to the creative brief is the next most critical factor. This encompasses both textual accuracy and visual interpretation. Within this tier, accuracy in text is paramount. This is evaluated in \textbf{point 3 (Typographical Precision)} and \textbf{point 4 (Visual Prompt Interpretation)}.
            \vspace{0.3em}
            \item \textbf{Priority 3: AI-Generated Artifacts.} Images that are technically sound and prompt-adherent are then judged on their level of polish and realism. The absence of distracting AI-specific rendering errors, such as anatomical distortions or illogical object blending, is the third most important factor. This is evaluated in \textbf{point 2 (AI Artifact and Realism Evaluation)}.
            \vspace{0.3em}
            \item \textbf{Priority 4: Standalone Aesthetic Appeal.} This is the final consideration, used primarily to differentiate between images that perform similarly in the top three tiers. It is a subjective measure of the image's artistic merit, including composition, lighting, and overall impact. This is evaluated in \textbf{point 5}.
        \end{itemize}

        \vspace{0.5em}
        \textbf{The Creative Brief (Prompt) is as follows:} \\
        ``\{creative\_brief\}''

        \vspace{0.5em}
        \textbf{Required Output:} \\
        Your final output must be structured as follows. Do not include any conversational introductions or summaries outside of this defined structure.
        First, provide your structured analysis for each of the \{num\_images\} images. Use a clear heading for each image (e.g., \textbf{Image 1 Analysis}). Within each analysis, you must use the five numbered subheadings in the specified order.
        Following the analysis of all \{num\_images\} images, provide the final ranking in a single, clean JSON block. The JSON block must be the very last thing in your response and contain only the ranking.

        \vspace{0.5em}
        \textbf{(Example of the final part of the output)} \\
        ... \\
        \textbf{Image \{num\_images\} Analysis} \\
        \textbf{1. Fundamental Image Integrity:} \\
        \textnormal{[Your analysis here.]} \\
        ... \\
        \textbf{5. Standalone Artistic Evaluation:} \\
        \textnormal{[Your analysis here.]} \\

        \vspace{0.5em}
\begin{verbatim}
```json
{
"rank": {ranking_example}
}
```
\end{verbatim}
        \vspace{0.5em}
        *Replace the numbers with the image numbers (1-\{num\_images\}) in order from best to worst.*
    \end{example}
    \caption{\textbf{Prompt used for Multi-Round Ranking Data Collection (Part 2: Synthesis and Output).}}
    \label{prompt:ranking-collection-p2}
\end{figure*}

\begin{figure*}[ht]
    \setlength{\abovecaptionskip}{0.1cm} 
    \setlength{\belowcaptionskip}{0cm}

    \begin{example}{MLLM Preference Assessment Prompt}
        \small 
        \textbf{System Instruction:} You are an AI Image Quality Analyst. Your task is to evaluate a pair of AI-generated images based on a creative brief and determine their data quality.
        
        \vspace{0.5em}
        \textbf{Evaluation Criteria:} \\
        Analyze each image based on its:
        \begin{enumerate}[leftmargin=*, nosep, label=\arabic*.] 
            \item \textbf{Fundamental Image Integrity:} Technical quality (clarity, exposure, no major flaws).
            \item \textbf{AI Artifacts and Realism:} Presence of AI artifacts, anatomical correctness, overall realism.
            \item \textbf{Typographical Precision:} (If text is present) Accuracy of content, font, style, and layout.
            \item \textbf{Visual Prompt Interpretation:} Faithfulness to the creative brief's theme, composition, style, and elements.
            \item \textbf{Standalone Artistic Quality:} Aesthetic appeal, composition, lighting, and emotional impact.
        \end{enumerate}
        
        \vspace{0.5em}
        \textbf{Decision Task:} \\
        Based on a holistic evaluation of the criteria above, you must categorize the image pair into one of the following four options.
        
        \vspace{0.5em}
        \textbf{Output Options:}
        \begin{itemize}[leftmargin=*, nosep]
            \item \textbf{``Image 1''}: If Image 1 is clearly better than Image 2.
            \item \textbf{``Image 2''}: If Image 2 is clearly better than Image 1.
            \item \textbf{``Tie''}: If both images are good and of comparable quality.
            \item \textbf{``Both are bad''}: If both images significantly fail the prompt's core requirements. This includes, but is not limited to: major errors in text or subject matter, missing key content, severe image quality issues, or extremely low aesthetic value.
        \end{itemize}
        
        \vspace{0.5em}
        \textbf{The Creative Brief (Prompt):}
        \begin{quote}
            ``\{creative\_brief\}''
        \end{quote}
        
        \textbf{Required Output Format:} \\
        Your final output must contain \textbf{only one} of the four decision strings and nothing else. Do not provide analysis, justifications, or any other conversational text.
        
        \vspace{0.5em}
        \textbf{Example:} \\
        Image 1
    \end{example}
    \caption{\textbf{Prompt used for MLLM Pairwise Preference Verification.} This prompt guides the model to evaluate image pairs across five dimensions, specifically filtering out low-quality samples via the ``Both are bad'' option.}
    \label{prompt:3-1-pairwise-preference}
\end{figure*}

\end{document}